\documentclass[12pt,preprint]{aastex}

\begin{document} 

\def\U        {{$U$ }}
\def\B        {{$B$ }}
\def\V        {{$V$ }}
\def\R        {{$R$ }}
\def\I        {{$I$ }}
\def\Uband    {{$U_{3600}$} }
\def\Bband    {{$B_{450}$} }
\def\Bj       {{$B_{j}$} }
\def\Vband    {{$V_{606}$} }
\def\Iband    {{$I_{814}$} }
\def\cge      {{$_ >\atop{^\sim}$}}
\def\cle      {{$_ <\atop{^\sim}$}}
\def\etal     {{et\thinspace al. }}
\def\eg       {{e.\thinspace g. }}
\def\cg       {{c.\thinspace g. }}
\def\ie       {{i.\thinspace e. }}
\def\SN       {{$S/N$} }
\def\Ho       {{$H_{0}$} }
\def\qo       {{$q_{0}$} }
\def\Lstar    {{$L^{*}$} }
\def\magarc   {{\ mag\ arcsec$^{-2}$} }

\title{Automated Galaxy Morphology: A Fourier Approach}

\author{S. C. Odewahn$^{1}$, S. H. Cohen$^{1}$, R. A. Windhorst$^{1}$, \& Ninan Sajeeth Philip$^{2}$ }
\affil{$^{1}$Dept. of Physics and Astronomy, Arizona State University} 
\affil{$^{2}$Dept. of Physics, Cochin University of Science and Technology, Kochi-682 022, India} 

\begin{center}
{\bf Date of this version: {\large \today} } 
\end{center}

\begin{abstract} 
 We use automated surface photometry and pattern classification 
techniques to morphologically classify galaxies. The two-dimensional 
light distribution of a galaxy is reconstructed using Fourier 
series fits to azimuthal profiles computed in concentric elliptical 
annuli centered on the galaxy. 
Both the phase and amplitude of each Fourier component 
have been studied as a function of radial bin number for a large 
collection of galaxy images using principal component analysis. 
We find that up to 90\% of the variance in many of these Fourier 
profiles may be characterized in as few as 3 principal components 
and their use substantially reduces the dimensionality of the 
classification problem. We use supervised learning methods in the 
form of artificial neural networks to train galaxy classifiers that 
detect morphological bars at the 85-90\% confidence level and 
can identify the Hubble type with a $1\sigma$ scatter of 1.5 
steps on the 16-step stage axis of the revised Hubble system. Finally,
we systematically characterize the adverse effects of decreasing 
resolution and \SN on the quality of morphological information 
predicted by these classifiers. 
\end{abstract} 

{\it Subject headings:} galaxies: automated morphological classification 


\section{INTRODUCTION} 
   One of the earliest galaxy classification schemes, 
discussed by Hubble (1926), was based on the visual appearance of 
two-dimensional images. This and other early schemes, based largely 
on photographic images in roughly the \B bandpass, used global 
image properties such as the visually perceived bulge-to-disk 
ratio and the degree of azimuthal surface brightness symmetry as major 
classification criteria. Even these early systems discussed 
the presence of spiral structure and the overall 
characteristics of the spiral arms: grand design of high or 
low pitch angle verses patchy or multiple arms (later termed 
flocculent spirals). The early Hubble system delineated spirals 
into a parallel sequence of barred and unbarred disk galaxies 
in a two-dimensional system. A refinement of this approach by  
de Vaucouleurs (1959), established a three-dimensional 
classification volume with the major axis being the Hubble 
stage (E,S,I) and the two other axes representing family (barred 
or unbarred) and variety (ringed or non-ringed). The revised 
Hubble system (hereafter referred to as the RHS) was designed to 
describe a continuum of morphological properties. Early workers 
in the field strove to systematize the degree of visual 
structures we see in galaxies (bulge/disk ratio, spiral arms, 
bars, rings) in an effort to understand the physical processes 
that formed these different galaxies. 

A useful classification system must relate members of different
classes to well understood general properties of the objects 
being systematically classified. It has long been known 
(de Vaucouleurs 1977, Buta \etal 1994, and Roberts and Haynes 1994)
that the stage axis of the Hubble sequence produces smooth, 
strong correlations with well known global properties: color,
surface brightness, maximum rotational velocity, and gas content.
These measured quantities are linked directly 
to very important physical properties: stellar population fraction, 
the surface density of stars, total mass of the system, and the 
rate of conversion of gas to stars. As discussed by 
Odewahn and Aldering (1995), such correlations may be smooth 
and significant in a mean sense, however the degree of 
accidental scatter in these relations is quite large. In 
most cases this scatter is much larger than that introduced 
by measurement error and is clearly cosmic in origin.  
The family and variety estimates of the  RHS 
produce much less significant correlations with such properties. 
Buta and Combes (1995), in an excellent review of these morphological 
features, make clear that family and variety convey information 
about the specific dynamical 
properties of disk systems: the degree of differential rotation,
the presence of certain families of resonant orbits among the 
stellar component of the disk, and the response of the viscous 
disk components (the gas) to the global and local properties 
of the galaxy potential well.

 Galaxy classification systems are generally built on the visual 
appearance in an image, and hence data sets based on these systems
can be gathered from imaging surveys alone. The mean correlations
described above are then used to draw inferences about important 
physical processes describing galaxies via these catalogs. In 
other words, morphological classification provided a "cheap" and 
direct means of acquiring large statistical samples describing 
galaxies. In some respects, the qualitative nature of this classification 
approach, and the dearth of recognized experts who classified in 
some established system gave the approach a less than deserved 
poor reputation. However, even recent comparisons among the 
estimates of experienced human classifiers (Naim \etal 1995) show 
a reliable level of repeatability is achieved.  
With an expansive increase in the quantity and quality of galaxy 
images in the last decade data via HST and a host of ground-base 
imaging surveys, interest has returned to the field of morphological 
classification (Driver \etal 1995a,b; Glazebrook \etal 1995; Abraham \etal 1996;
Odewahn \etal 1996).  Many workers have developed automated image 
analysis systems that provide quantitative estimates of galaxy types 
(Okamura \etal 1984, Whitmore 1984, Spiekermann 1992, Storrie-Lombardi 
\etal 1992, Abraham \etal 1994, Odewahn 1995, Han 1995, Odewahn \etal 1997). 
In one the most physically meaningful approaches, Rix and Zaritsky (1995)  
showed in a small sample of spiral galaxies that kinematic distortions 
in the global velocity field of a galaxy can be 
linked to global asymmetry measurements in the light distribution.   
A discussion of novel feature spaces for identifying peculiar 
galaxies is presented by Naim \etal (1997). 
All of these studies have produced a wealth of information about the 
systematic properties of galaxies, particularly in the area 
of comparing the low and high redshift Universe (Abraham \etal 1996, 
Odewahn \etal 1996).  However, 
it must be conceded that none of these classification systems 
are truly morphological in nature. In general, each method 
uses one or more two-dimensional parameter spaces based 
on global image properties 
and produces a type estimate via mean correlations like those 
discussed above using some multivariate pattern classification 
approach (\eg linear parameter space divisions, 
principal component analysis, decision tree, artificial neural 
network). 

In this paper we describe a truly morphological approach to 
galaxy classification based on the Fourier reconstruction of 
galaxy images and the subsequent pattern analysis based on 
the amplitude and phase angle of the Fourier components used. 
This method can quantitatively detect the presence of 
spiral arms and bars. In addition, it provides a systematic 
means of describing the degree of large-scale 
global asymmetry in a galaxy, a property known to correlate loosely
with Hubble stage, but one which is also strongly linked to 
important physical events such as merging and tidal interaction. 

In Section 2 we describe our machine-automated classification method. 
In Section 3 we compare independent sets of visual estimates of 
morphological types in the RHS that establish a well understood 
set of training/testing cases.  These date are comprised of 
a local sample from ground-based imaging and a distant 
sample collected from the HST archive. 
In Section 4 we demonstrate a practical application using these 
data to train and assess the performance of several Fourier-based 
galaxy classifiers. We discuss the role of systematic and 
accidental errors in its use caused by varying image resolution 
and \SN.  In Section 5 we summarize the results and lay the groundwork for 
future morphological studies of distant galaxy populations.

\section{CLASSIFICATION METHODOLOGY}

The two-dimensional luminosity distribution observed in most galaxies
most often displays a high degree of azimuthal symmetry. This 
justifies the use of one-dimensional radial surface brightness profiles
in describing the radial stellar surface density distribution. Such 
profiles are then decomposed into constituent photometric components
(de Vaucouleurs 1958, de Vaucouleurs and Simien 1986, Kent 1986,
de Jong and van der Kruit 1994).  This profile is extracted from 
the galaxy image in many ways, but most often by computing mean flux 
density in elliptical annuli having the shape and orientation of 
the faint surface brightness, outer isophotes. Parameterizing this 
profile, \cg with a concentration index, will produce a measurement 
that correlates well with bulge-to-disk ratio and hence with Hubble 
type. Local departures from this overall radial symmetry 
in the form of high spatial frequency components (arms, bars, rings) 
form the basis for additional criteria to visual galaxy classification 
systems like the RHS.  

 A method which quantitatively describes departures 
from azimuthal symmetry in galaxy images would seem to be a rich 
starting place for developing a machine automated classification 
system, either of the supervised learning variety 
(Odewahn \etal 1992, Weir \etal 1995) or the unsupervised 
variety (Mahonen \etal 1995). Towards this end, we have   
adapted a moments-based image analysis technique (Odewahn 1989) 
to fit Fourier components to fixed-grid azimuthal profiles in 
elliptical annuli to reconstruct galaxy images.
One may think of this step as an optimized data compression 
technique for reducing the dimensionality of our ultimate 
pattern classification problem: the recognition and characterization 
of morphological features in galaxies. 

  The technique of Fourier decomposition presented here is used to 
quantify the two dimensional luminosity distributions of galaxies. 
Similar approaches have been used to study the shapes of isophotes 
in elliptical galaxies by Bender and Mollenhoff (1987), and 
applications to the luminosity distribution are described 
by Lauer (1985), Buta (1987), and Ohta (1990).  With this 
methodology, we can quantify the amplitude and phase of the bar 
and spiral arm components in a galaxy image. This technique is very useful 
for modeling the bar luminosity distribution for systems in which
a simple elliptical bar model was not sufficient, as was first 
demonstrated by Elmegreen and Elmegreen (1985) in describing the bar 
luminosity distributions exhibited by galaxies over a range of 
Hubble types. 

\subsection{Pattern Classification Using Artificial Neural Networks} 
Artificial neural networks (ANN's) are systems of weight vectors, whose
component values are established through various machine learning algorithms.
These systems receive information in the form of a vector (the input pattern) 
and produce as output a numerical pattern encoding a classification. 
They were designed to simulate groups of biological neurons and 
their interconnections and to mimic the ability of such systems 
to learn and generalize.  Discussion of the development and 
practical application of neural networks is well presented in 
the literature (see McClellan and Rumelhardt 1988). 
In astronomical applications, this technique has been applied with
considerable success to the problem of star-galaxy separation by 
Odewahn \etal (1992). 
A neural network classifier was developed Storrie-Lombardi \etal (1992) 
to assign galaxy types on the basis of the photometric parameters 
supplied by the ESO-LV catalog of Lauberts and Valentijn (1989).  
Applications of this method to large surveys of galaxies using
photographic Schmidt plate material are discussed in 
Odewahn (1995) and Naim \etal (1995). More recent discussions of 
star-galaxy separation using ANN's are presented by 
Andreon \etal (2000), Mahonen \& Frantti (2000), and 
and Philip \etal (2000).  An extremely thorough discussion 
of galaxy classification with ANN methods is given by Bazell (2000).   

The neural network literature quoted above contains a wealth of
information on the theoretical development of neural networks and
the various methods used to establish the network weight values.
We summarize here only the practical aspects of the ANN operation
and the basic equations for applying a feed-forward network. 
The input information, in our case the principal components 
formed from the Fourier profiles discussed in Section 3.2, is presented
to the network through a set of input nodes, referred to as the input layer.
Each input node is ``connected'' via an information pathway to nodes
in a second layer referred to as the first hidden layer. In a
repeating fashion, one may construct a network using any number of such
hidden layers. For the backpropagation networks discussed here, we
have used ANN's with two hidden layers. The final layer of node positions,
referred to as the output layer, will contain the numerical output of the
network encoding the classification. This system of nodes and 
interconnections is analogous to the system of neurons and 
synapses of the brain. Information is conveyed along each 
connection, processed at each node site, and passed
along to nodes further ``upstream'' in the network. This
is referred to as a feed-forward network. The information processing
at each node site is performed by combining all input numerical information
from upstream nodes, $a_{pj}$, in a weighted average of the form:      
\begin{equation}
\beta_{i} = \sum_{j} w_{ij} a_{pj} + b_{i}    \label{eq:eq1}
\end{equation}
 The $j$ subscript refers to a summation of all nodes in the previous
layer of nodes and the $i$ subscript refers to the node position in the
present layer, i.e. the node for which we are computing an output. Each time
the ANN produces an output pattern, it has done so by processing information
from the input layer. This input information is referred to as the input
pattern, and we use the $p$ subscript to indicate input pattern. Hence,
if there are 5 nodes in the previous layer ($j$), each node ($i$) in the
current layer will contain 5 weight values ($w_{ij}$) and a constant
term, $b_{i}$, which is referred to as the bias. The final nodal output
is computed via the activation function, which in the case of this
work is a sigmoidal function of the form:
\begin{equation}
a_{pi} = \frac{1}{1+\exp^{-\beta_{i}}}    \label{eq:eq2}
\end{equation}
Hence, the information passed from node $i$ in the current layer of
consideration, when the network was presented with input pattern $p$, is     
denoted as $a_{pi}$. This numerical value is subsequently passed to the
forward layer along the connection lines for further network processing. The
activation values computed in the output layer form the numerical
output of the ANN and serve to encode the classification for input
pattern, $p$.
                
 In order to solve for the weight and bias values of equation 1 for
all nodes, one requires a set of input patterns for which the correct
classification is known. This set of examples is used in an iterative
fashion to establish weight values using a gradient descent algorithm
known as backpropagation. In brief, backpropagation training is performed
by initially assigning random values to the $w_{ij}$ and $b_{i}$ terms
in all nodes. Each time a training pattern is presented to the ANN, the
activation for each node, $a_{pi}$, is computed. After the output layer
is computed, we go backwards through the network and compute an error
term which measures of the change in the network output produced by
an incremental change in the node weight values. Based on these 
error terms, all network weights are updated and the process 
continues for each set of training patterns. As discussed in the 
next section, precautions must be taken in this process to avoid 
problems from over-training (\ie simply memorizing the input 
pattern set as opposed to generalizing the problem).  

  The goal of the present work is to develop a system that 
automatically detects the presence of specific types of morphological 
structures in galaxy images. This is clearly a more complex problem 
compared to the use of a few global photometric parameters  to 
predict types, and so we desired to use an additional method of 
neural network classification to confirm the 
results of our classical backpropagation approach. 
Philip \etal (2000) develop a neural network based on 
Bayes' principal that assumes the clustering of attribute 
values (the input parameters). The method considers the 
error produced by each training pattern presented to the 
network, and weights are updated on the basis of the 
Bayesian probability associated with each attribute for 
a given training pattern. In this approach, the probability 
density of identical attribute values flattens out  while 
attributes showing large differences from the mean get 
boosted in importance. This is referred to as a 
difference boosting neural network (DBNN).  We have applied 
this new approach as a check on our backpropagation 
network results, however this method has the added 
advantage that it trains to convergence in a much faster 
time than is required by classical backpropagation methods.

\subsection{Initial Image Reconstruction} 
In this paper we develop a technique that is a pre-processing 
step for producing input to a supervised classifier in the 
form of an artificial neural network (Odewahn 1997). In 
this approach, the radial surface brightness profile of a galaxy  
is computed in elliptical annuli centered on the galaxy center. 
The rational here, valid in most cases, is that we are isolating the 
disk component of the galaxy and hence establishing a way 
of defining the equatorial plane of the system. 
The position, shape and orientation of these annuli are 
determined using classical isophotal ellipse fitting 
techniques. Within each annulus, we compute the run 
of flux density with position angle, $\theta$, in the 
equatorial plane of the galaxy, to form the azimuthal 
surface brightness profile.  To describe each azimuthal 
profile, and reduce the dimensionality of our classification 
problem, each azimuthal profile is modeled by the following 
Fourier series:
\begin{equation}
 I_{o}(r,\theta) = I_{\circ} + \Sigma I_{mc}(r) \cos m\theta + \Sigma I_{ms}(r) \sin m\theta    \label{eq:fourier0}
\end{equation}
For computational efficiency, the Fourier terms are computed for 
each azimuthal profile using the following moment relations :
\begin{equation}
 I_{o}(r) = < I(r,\theta) >                   \label{eq:fourier1}
\end{equation}
\begin{equation}
 I_{mc}(r) = 2 < I(r,\theta) \cos m\theta >    \label{eq:fourier2}
\end{equation}
\begin{equation}
 I_{ms}(r) = 2 < I(r,\theta) \sin m\theta >    \label{eq:fourier3}
\end{equation}
where $\theta$ is the angle (in the equatorial plane of the galaxy) with
respect to the photometric major axis, and m is an integer. Hence, if 
we assume galaxy  disks to be thin and suffer no internal extinction, 
this approach crudely removes inclination effects and allows us to 
compare galaxy images independent of orientation. The Fourier series 
in this paper were computed up to m=5. Fourier amplitudes, 
describing the relative amplitude of each component are computed 
using the following expression :
\begin{equation}
 A_{m}(r) = \sqrt{ I_{mc}(r)^{2} + I_{ms}(r)^{2} } / I_{o}(r)  \label{eq:fourier4}
\end{equation}
Finally, a phase angle describing the angular position of the 
peak signal contributed by each component can be computed. 
We compute the phase angle of the $m\theta$ component as: 
\begin{equation}
\theta_{\circ} = (1/m) \tan^{-1}(I_{ms}/I_{mc})  \label{eq:fourier5}
\end{equation}

\begin{figure}
 \epsscale{1.0}
 \plotone{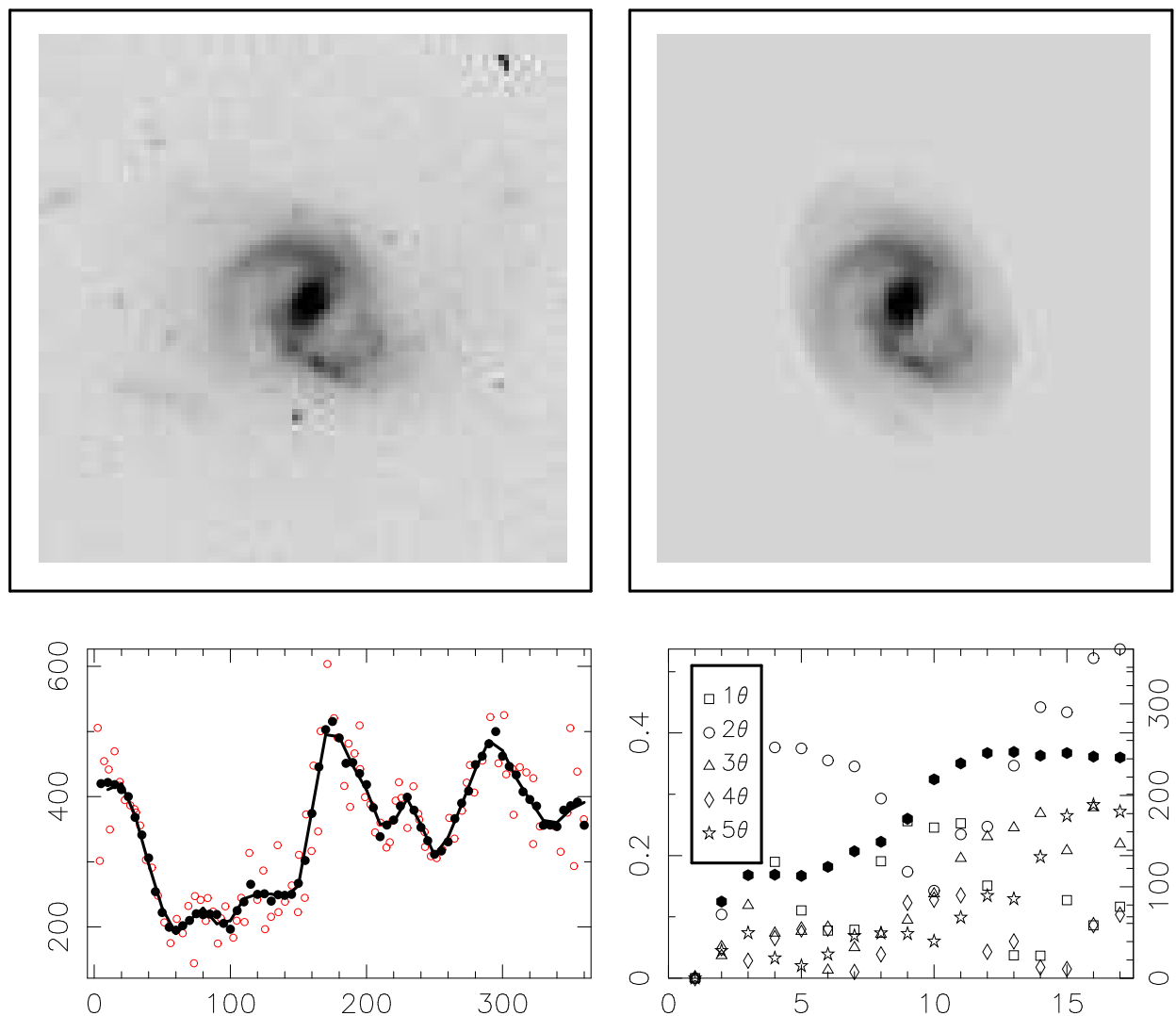}
\figcaption[f1.eps]{\sf \footnotesize The Fourier image-modeling method 
         described in the text applied to a galaxy 
         observed with HST in the B-Band Parallel Survey 
         (BBPS, Cohen \etal 2001). The original \I image is shown 
         in the upper-left panel, and the Fourier model image 
         is shown in the upper-right. In the lower-left panel we
         plot the azimuthal profile for the annulus over-plotted 
         in the galaxy image. Open circles are individual pixel 
         values, solid points represent the smooth points, and the 
         solid line represents the Fourier series fitted to this 
         profile. In the lower-right panel we plot the five Fourier 
         amplitudes as a function of radial annulus number using 
         open symbols (left numeric scale). Finally, the solid points 
         represent the phase angle, measured in degrees in the equatorial 
         plane, of the 2$\theta$ component (right numeric scale). 
         \label{model_sample_1}}
\end{figure}

\begin{figure}
 \epsscale{1.0}
 \plotone{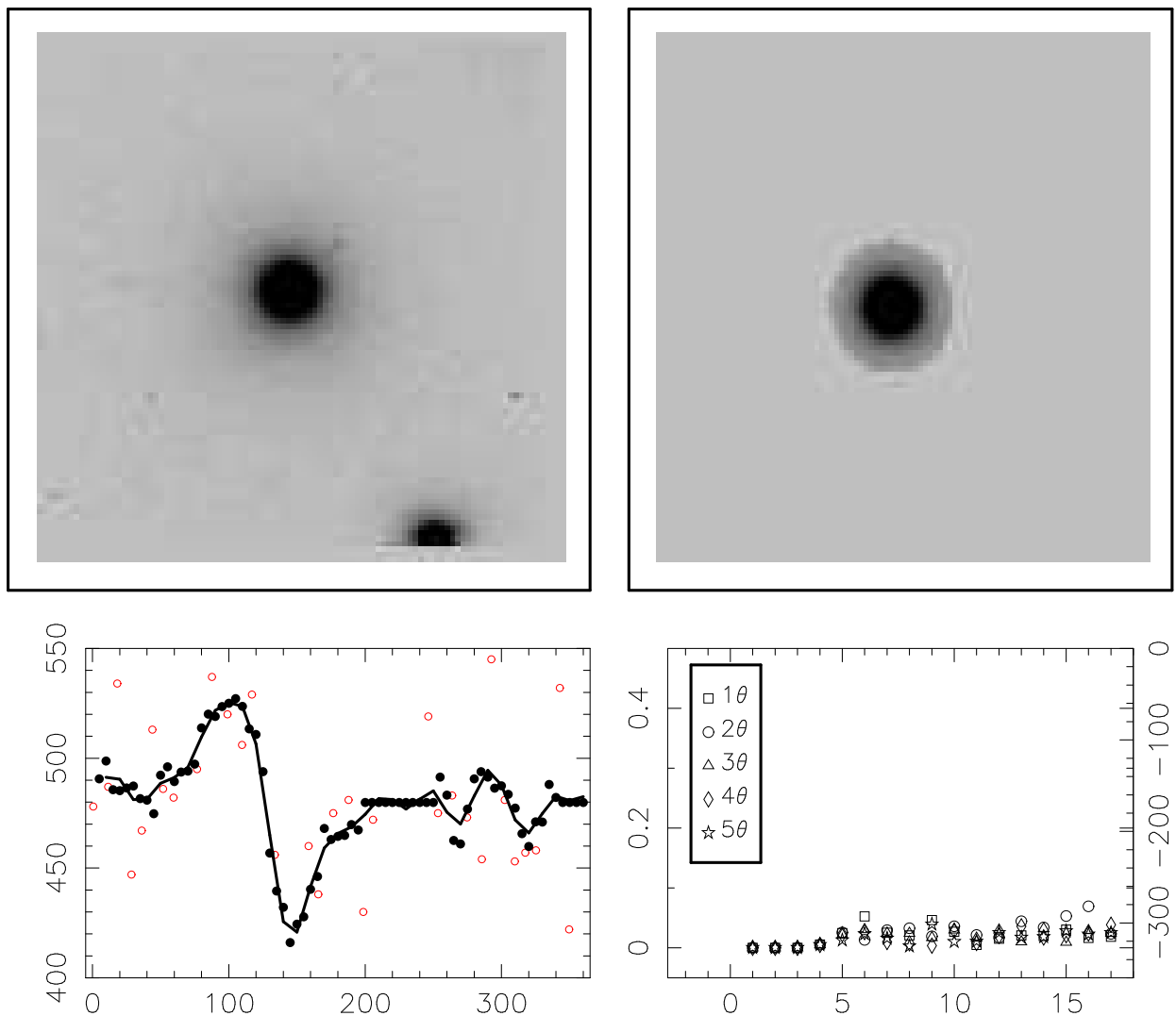}
\figcaption[f2.eps]{\sf \footnotesize The same style of figure as in 
         Figure 1, except we now measure an early-type galaxy from 
         the BBP. In this case, there is little 
         high spatial frequency image structure and only very small 
         departures from radial symmetry. Hence, the Fourier amplitudes,
         plotted to the same scale as in Figure 1, are very small. 
         Additionally, no phase angles are plotted since the $2\theta$ 
         Fourier amplitudes were below the \SN limit imposed for 
         computing these angles. \label{model_sample_2}}
\end{figure}

A practical application of this method to a barred spiral galaxy imaged 
with HST is shown in Figure~\ref{model_sample_1}.  Another example for 
an elliptical galaxy image which contains no high spatial frequency image 
structure is shown in Figure~\ref{model_sample_2}. Through experimentation 
with both HST- and ground-based images, we have determined that 
using 17 elliptical annuli and up to $m=5$ in Equation \ref{eq:fourier0} 
consistently reproduces the basic morphological features 
of most galaxy images. A considerable amount of experimentation 
went into determining how one should establish the radial bin 
intervals in this procedure. A series of tests were made using 
unequal linear bins that overlapped and logarithmic binning interval. 
We experimented with normalizing the bin radius by the effective 
radius, $r_{e}$, so that all galaxies could be compared on more uniform 
spatial scale. This approach proved problematic since it produced  
profile scales that were highly compressed for early-type systems,
a highly expanded in late-type systems. In such cases, the inner regions 
if the late-type galaxies, where much of the important morphological 
structure is present, was under-resolved. Additionally, we found that 
measurement of the effective radius of galaxies in low resolution, 
low \SN HST images can be systematically low, and using $r_{e}$ in 
this manner will introduce a substantial bias when comparing local and 
distant samples. In the end, we found that using equal size radial 
bins extending to an outer, low surface brightness optical isophote 
produce the most robust results for comparing morphological properties 
among galaxy images.  

  It should be noted that one can justify the 
use of higher order Fourier components for cases of high resolution
galaxy images. In experiments with some of the deep Palomar 60$''$ 
images described in Section 3, we found that the bars of many 
galaxies have a strong contrast relative to the disk, and  high 
spatial frequency components (up to $12 \theta$) are required 
to reconstruct such a sharp feature. Incorporating so many spatial
frequencies in a general image classifier is impractical, but one 
might hope to use some type of information about high frequency 
structure: if the inclusion of such components in the model 
significantly improves the model fit, then this is important 
information. Perhaps a good number of barred galaxies can be 
identified if we simply search for images that require high 
order Fourier components in the inner annuli to build a good 
fit model.    

\subsection{Dimensionality Reduction using Radial Trends} 
  The Fourier-reconstructed images described above are 
comprised of radially-dependent sets of Fourier amplitudes and 
phase angles. Using 17 annuli, 6 Fourier amplitudes (including 
the $m=0$ term) and the corresponding phase angles to parameterize 
each image, we thus describe each galaxy with a 221 element 
classification vector. Some human classifiers have insisted that 
at least 10,000 resolution elements are required in a galaxy 
image that is to be morphologically classified (de Vaucouleurs, private
communication). More liberally, we might consider that one should 
have at least a $50\times50$ image of a galaxy, or 
2500 pixels, to convey morphological information. 
Hence, using the Fourier description of a galaxy reduces the number 
of parameters describing the image information content by a factor 
of more than ten compared to that used by a human classifier.  
An input classification vector 
with over 100 elements is rather large for most pattern 
classifiers trained in the presence of noise.
To distill our image parameterization further, we have 
chosen to characterize the radial properties of each Fourier 
component. In other words, just as we often describe the 
radial surface brightness profile of a galaxy using several 
model parameters, so too can we describe the radial trends 
in each set of amplitude and phase estimates computed with 
equations ~\ref{eq:fourier4} and ~\ref{eq:fourier5}.

\begin{figure}
 \plottwo{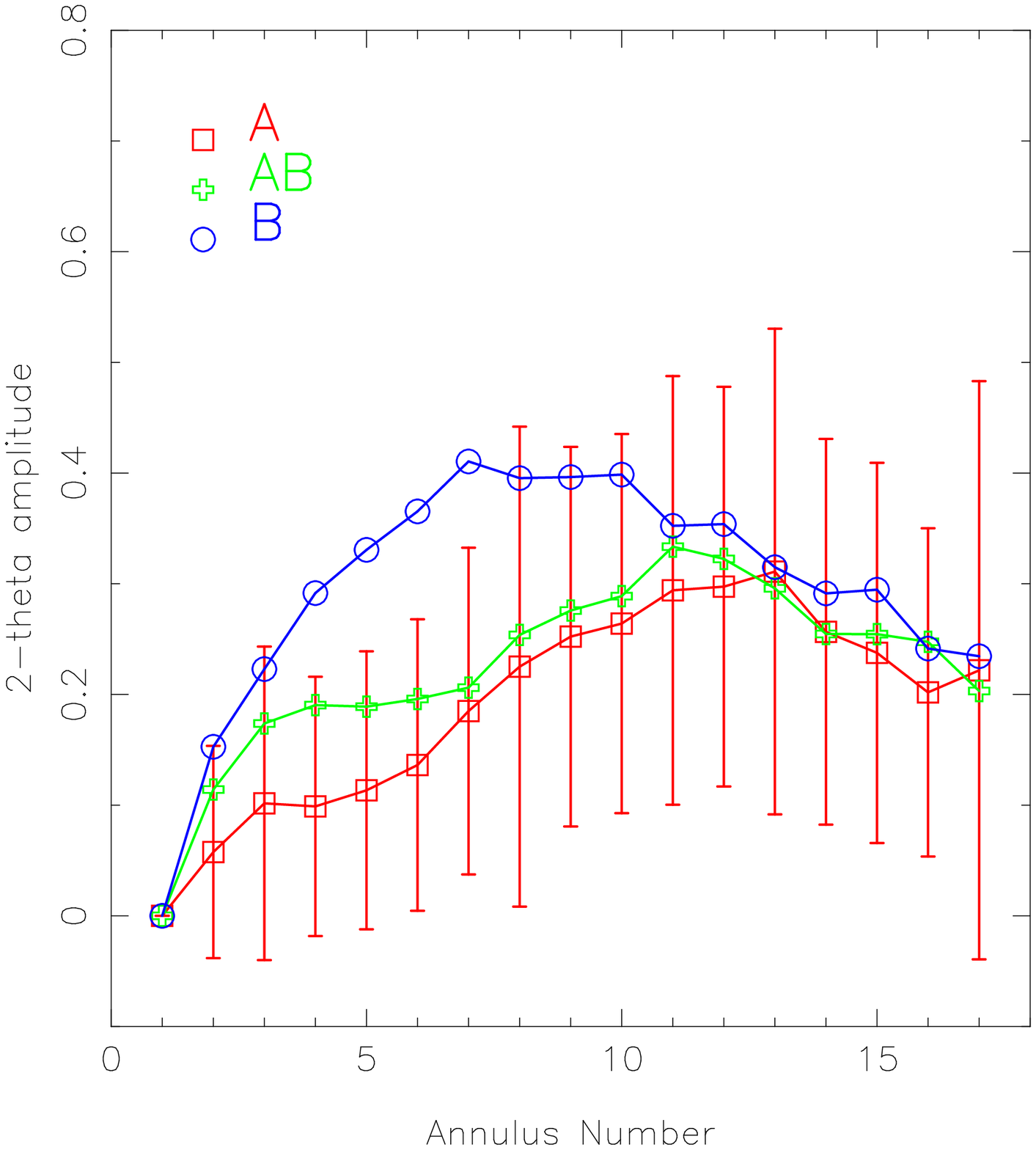}{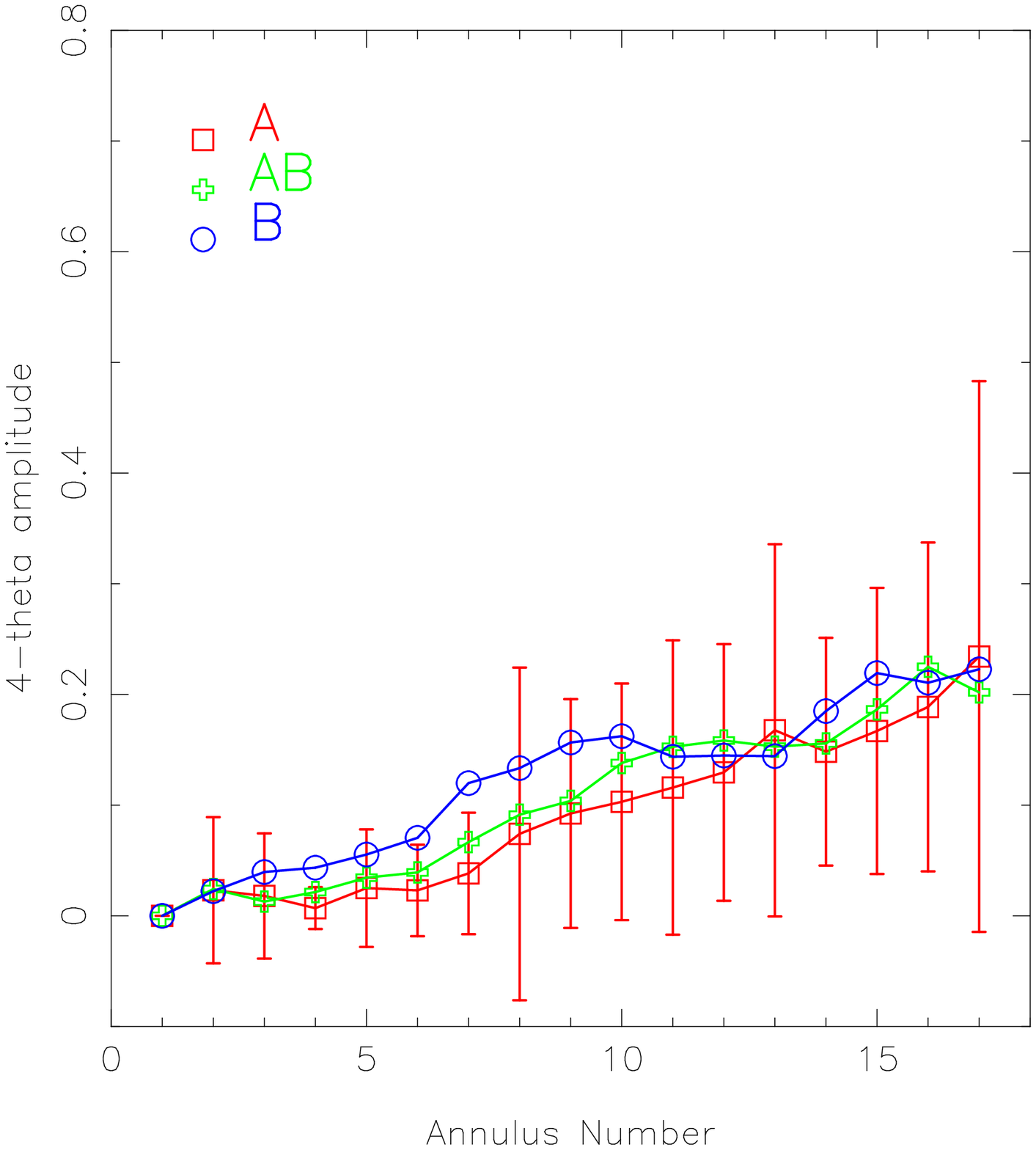}
\figcaption[f3a.eps,f3b.eps]{\sf \footnotesize Mean profiles of the 
         $2\theta$ (left) and $4\theta$
         (right) amplitudes computed for three sets of galaxies 
         divided by family class. We used  71 A, 73 AB, and 61 B 
         galaxies. The mean barred (B) galaxy  
         trends are well separated from unbarred (A) 
         galaxies for annuli numbers (on the x-axis) in
         the range 3 to 12. We plot the $1\sigma$ error bars 
         for the mean A points only. It is significant that 
         in both the $2\theta$ and $4\theta$ cases the 
         mean AB profiles lie between the A and B curves indicating 
         that these galaxies do indeed possess only weakly
         detected bar structures.  \label{MFPROF1}}
\end{figure}

\begin{figure}
 \plottwo{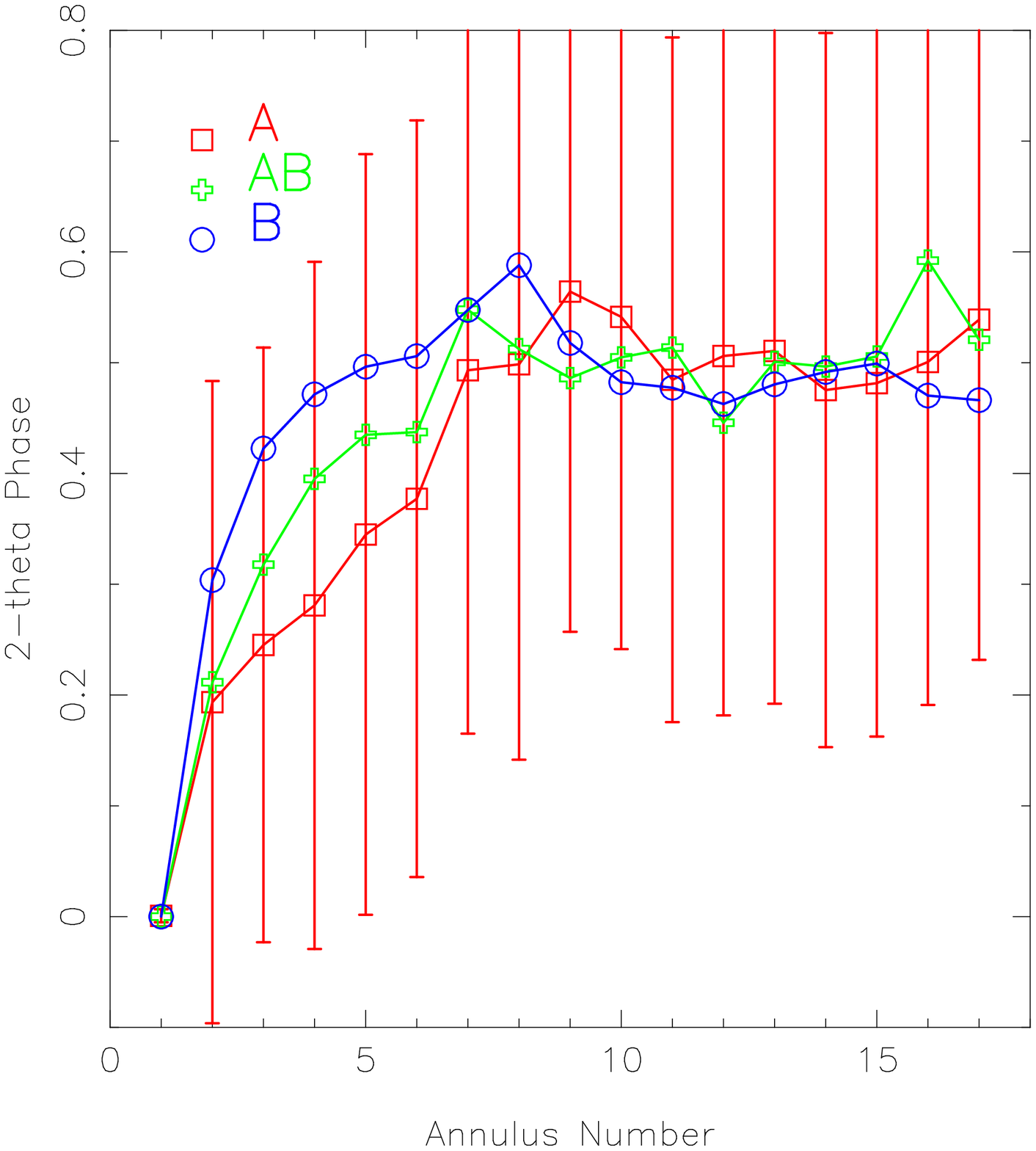}{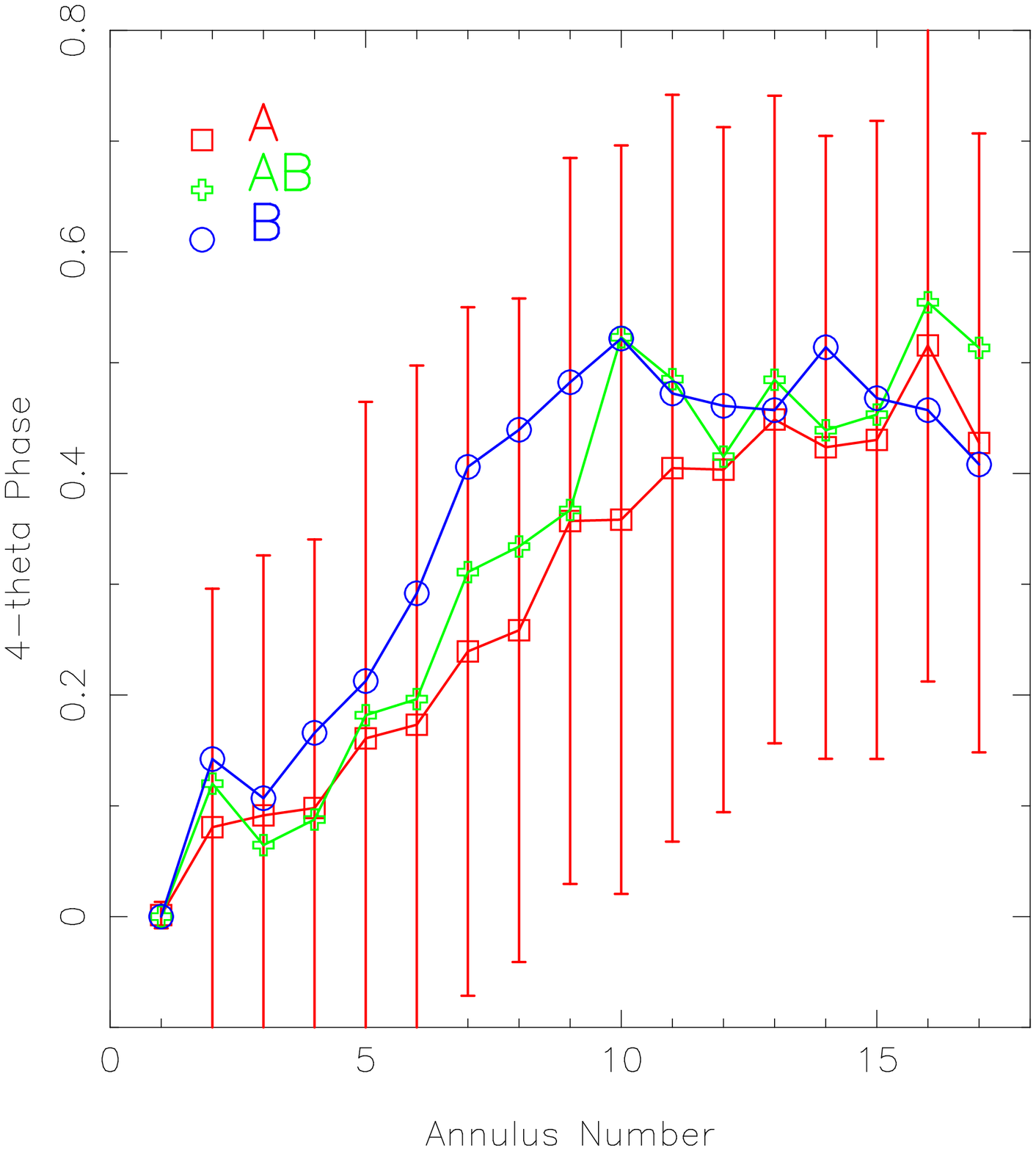}
\figcaption[f4a.eps,f4b.eps]{\sf \footnotesize Mean profiles of the 
         $2\theta$ (left) and $4\theta$ (right) 
         phase computed for the same samples used in 
         Figure~\ref{MFPROF1}.  Although amplitude information 
         clearly provides a more powerful means of detecting 
         the presence of a bar, phase information is needed to 
         differentiate between a purely linear structure like 
         a bar and a simple two-armed spiral pattern. \label{MFPROF2}} 
\end{figure}

 The mean radial trends in both the amplitude and phase 
of the coefficients in equations ~\ref{eq:fourier4} and 
~\ref{eq:fourier5} are shown in Figures ~\ref{MFPROF1}
and ~\ref{MFPROF2} for three types of galaxies. In 
these figures the mean phase or amplitude of a given Fourier 
component is plotted on the y-axis as a function of elliptical 
annulus number on the x-axis. We divided a 
sample of 196 galaxies from Section 3 into three family 
categories: A (unbarred system), 
AB (transitional systems), and B (barred systems) and computed 
the mean radial profiles for the amplitude and phase coefficients 
of the $2\theta$ and $4\theta$ coefficients. As noted by 
Elmegreen and Elmegreen (1985) and Ohta (1990), these are the 
dominant terms needed to reproduce the light distributions 
of most barred galaxies. In Figure~\ref{MFPROF1} we see a 
clear trend among the $2\theta$ and $4\theta$ amplitudes: 
barred systems have significant power in the inner rings,
AB systems have systematically lower power, and A galaxies 
have the lowest amount of $2\theta$ and $4\theta$ power 
in even the inner elliptical annuli. These mean profiles 
were computed with 71 A, 73 AB and 61 B galaxies taken 
from the analysis of Section 4.  The 1-$\sigma$ error of 
each point in the A-system profiles are shown in 
Figures ~\ref{MFPROF1} and ~\ref{MFPROF2}, and it is 
clear that a single point from the profile of one galaxy 
will have little discriminatory power in distinguishing 
bar class. However, if we combine the information of 
the points from annuli with ring numbers between 3 and 12, 
then we can expect to derive a robust estimate of whether 
a bar is present in the inner regions of a galaxy image. 
In this region of the reconstructed images, the mean 
Fourier profiles clearly delineate the presence of 
a bar.  In Figure~\ref{MFPROF2} we plot the corresponding mean 
profiles for the $2\theta$ and $4\theta$ phase angles. 
Although these profiles are less sensitive to bar presence, 
the phase terms will be important in distinguishing 
barred galaxies from purely spiral systems. In the case of a 
barred galaxy, the $2\theta$ and $4\theta$ phase angles remain 
fixed with ring number (radius), whereas with a two-armed spiral 
pattern we expect to see a systematic variation of $2\theta$ 
and $4\theta$ phase angles with radius since the flux density 
peaks are changing their position angle smoothly as we 
progress outward in the radial annuli. As can be seen 
in Figure~\ref{MFPROF2} we still observe a gradual progression 
in the trends of these phase angle profiles for the A, AB, 
and B galaxy sets.  To summarize, the radial $2\theta$ and 
$4\theta$ amplitude and phase angle profiles of a galaxy 
carry important information 
about the nature of the 2-D structure present in a galaxy 
image, and specifically allow us to robustly quantify the 
presence of a linear bar structure.   
 
\begin{figure}
 \epsscale{1.0}
 \plotone{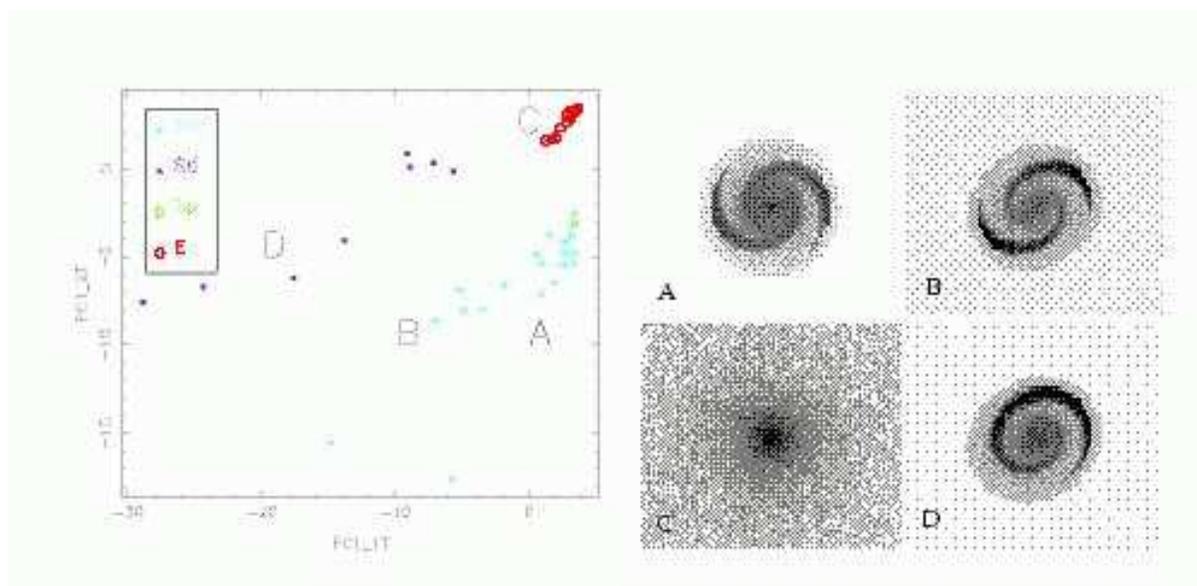}
\figcaption[f5.eps]{\sf \footnotesize (a) [Left] A parameter 
         space formed from the first principal component of the 
         1$\theta$ Fourier amplitude profile, PC\_1T, and the first 
         principal component of the 2$\theta$ Fourier amplitude 
         profile, PC\_2T. The large letter labels correspond to 
         the images in the right panel.  (b) [Right] Sample model 
         images produced by LMORPHO that were used to compute the 
         parameter space in the left-hand 
         panel. Objects in the top row (A,B) are idealized mid-type 
         spirals (with B inclined assuming an optically thin disk 
         model), object C (lower left) is a typical elliptical 
         model, and object D (lower right) is the one-armed spiral 
         morphology often seen in compact groups (see compact group 
         16 in the catalog of Hickson 1993). 
         We have used model galaxy images in this figure to clearly 
         demonstrate the power of the method, but tests with real 
         galaxy images have been found similarly effective. \label{pc_demo}}
\end{figure}

Following the innovative work of Han (1995), we performed  a
principal component analysis (PCA) of each set of Fourier profiles 
for our collections of galaxies that had reliable stage and 
family classifications. For most Fourier coefficients it 
was determined that 85\% to 90\% of the profile variance 
could be described with only 2 or 3 principal components. 
Hence, using the eigenvectors determined from this analysis, we 
can characterize the Fourier profile of a random galaxy using 
only 2 or 3 parameters. Should we use the amplitude and phase of
only the $2\theta$ and $4\theta$ components to characterize 
a galaxy, then we can parameterize the light distribution 
using 4 coefficients $\times$ 2 principal components $=$ 
8 parameters. Such an image vector constitutes a very 
reasonable size for training a supervised classifier with 
the 100-200 patterns (galaxies) available in this work.    
In Figures ~\ref{MFPROF1} and ~\ref{MFPROF2} we 
have divided the galaxies by family class, but these 
Fourier-based principal components may be used to identify 
stage-related image traits also. 
In Figure~\ref{pc_demo} we demonstrate our principal component method 
by showing PCA-based image parameter spaces that are symbol coded by 
stage and family class. 


\section{TRAINING AND TESTING SAMPLES}
 The development of a pattern classifier based on supervised learning, 
\ie a classifier that works in a predefined system, requires a data 
sample consisting of the input classification information (the image 
parameters) and the corresponding target patterns (the galaxy types).  
This sample is referred to as the training set, and a portion of it 
is held back from the actual supervised learning process (in our case 
backpropagation training) to be used as a test sample to make an 
unbiased judgement of classifier performance during and after 
training. As we discuss below, nearly all of the past morphology-based 
studies of high redshift galaxies have focused on the stage axis only.
Since our aim is to develop morphological classifiers that 
estimate the family and variety of galaxies based on Fourier amplitude 
strengths and phases, we have visually reclassified a large number 
of high \SN, well-resolved HST galaxy images in the RHS
to establish a statistically significant testing sample. Although 
the present sample is based on 35 medium-deep and deep WFPC2 fields, the 
number of galaxy images from this sample that are suitable for such 
classification is only 146. 
Hence, for the present study we incorporated the local galaxy 
samples from ground-based CCD observations by de Jong and van der Kruit 
(1995), hereafter J95, and Frei \etal (1996), hereafter F96, adding 
124 more galaxies to the sample. For both the local and 
distant HST samples we inter-compare our classification sets in order to 
objectively define a weighted mean training set whose systematic and 
accidental errors are well understood. From such a sample we may 
establish a fair "ground-truth" sample of objects for use in training 
and testing pattern classifiers. 

 A large collection of visual Type estimates in several bandpasses from 
WFPC2 deep fields were discussed in Odewahn \etal (1996) and 
Driver \etal (1995b). Similar visual classification sets have been 
assembled since that time for the Hubble flanking fields (Odewahn 
\etal 1997) and the BBPS fields (Cohen \etal 2001). To begin our 
assessment of classifier errors, we have analyzed a large number 
of past visual stage estimates made in the F814W bandpass images. 
We inter-compared all of the stage estimates common to the 
5 classifiers of Odewahn \etal (1996) and the DWG sample and 
derived linear transformations to convert all classifiers 
to a common system. The vast majority of classifications were 
from one classifier (1000 from Odewahn, hereafter referred to 
as SCO). It was decided that the Odewahn classes, which match well 
the RC3 mean Type system, would 
define the HST classification system. Following Odewahn and de Vaucouleurs 
(1993) we then use a residual analysis to determine the accidental 
scatter associated with each classifier, and hence a relative weight 
associated with each classification of an HST-observed galaxy. 
Finally, a catalog of 1306 galaxy types was assembled:
1000 with single Odewahn estimates with mean error of 
$\pm2.2$ type steps and 306 weighted mean types (109 having 5 
independent estimates) with mean error of $\pm1.2$ type steps.
The galaxies in this catalog have stage estimates from the    
the F814W images and objective errors estimated for each type. 
These data will hereafter be referred to as the HST1 set. 

  The purpose of our current work is to develop automated techniques 
that identify morphological structures, and with the exception of 
the bright BBP galaxy types estimates of SCO, the HST1 types are 
comprised only of estimates of the stage axis in the RHS 
(\ie E, S0, Sa, ..., Im). Such data sets are fine 
for the type classifiers discussed in Odewahn \etal (1996), however 
for more advanced classification experiments we desired 
a set of faint HST-observed galaxies with classifications in 
the RHS: stage, family and variety. To 
collect this set we extracted 4534 galaxy images in F814W 
from 39 different WFPC2 fields.  The 
majority of these images were contributed by moderate
depth WFPC2 parallel fields, but the deep 53W002 and HDF-N 
fields were also used.  From this sample we established, purely 
through visual inspection, the group of 146 galaxy images having 
sufficient \SN and resolution to allow good morphological classification. 
As a second sample, we binned 125 ground-based CCD images of 
F96 (in Gunn g) and J95 (in Johnson B) to a resolution similar 
to that of the HST images.   

  The manipulation of all galaxy images used in this work, whether 
for the purpose of human visual classification or for automated 
image analysis was performed with a variant of the MORPHO package 
discussed in Odewahn \etal (1997). A new version of this  
galaxy morphological classification software system called LMORPHO, 
which is optimized for use under the linux operating system, provides 
an image database manager that allows users to select galaxies 
for analysis based on morphological traits, photometric properties 
or many other meta-data properties (\cg bandpass, image size, or 
pixel scale).  Beyond the image database manager of LMORPHO, the 
system provides tools performing automated galaxy surface photometry, 
basic astronomical image processing, and a variety of multivariate
statistical analysis and pattern classification tasks. 

\begin{figure}
 \epsscale{1.0}
 \plotone{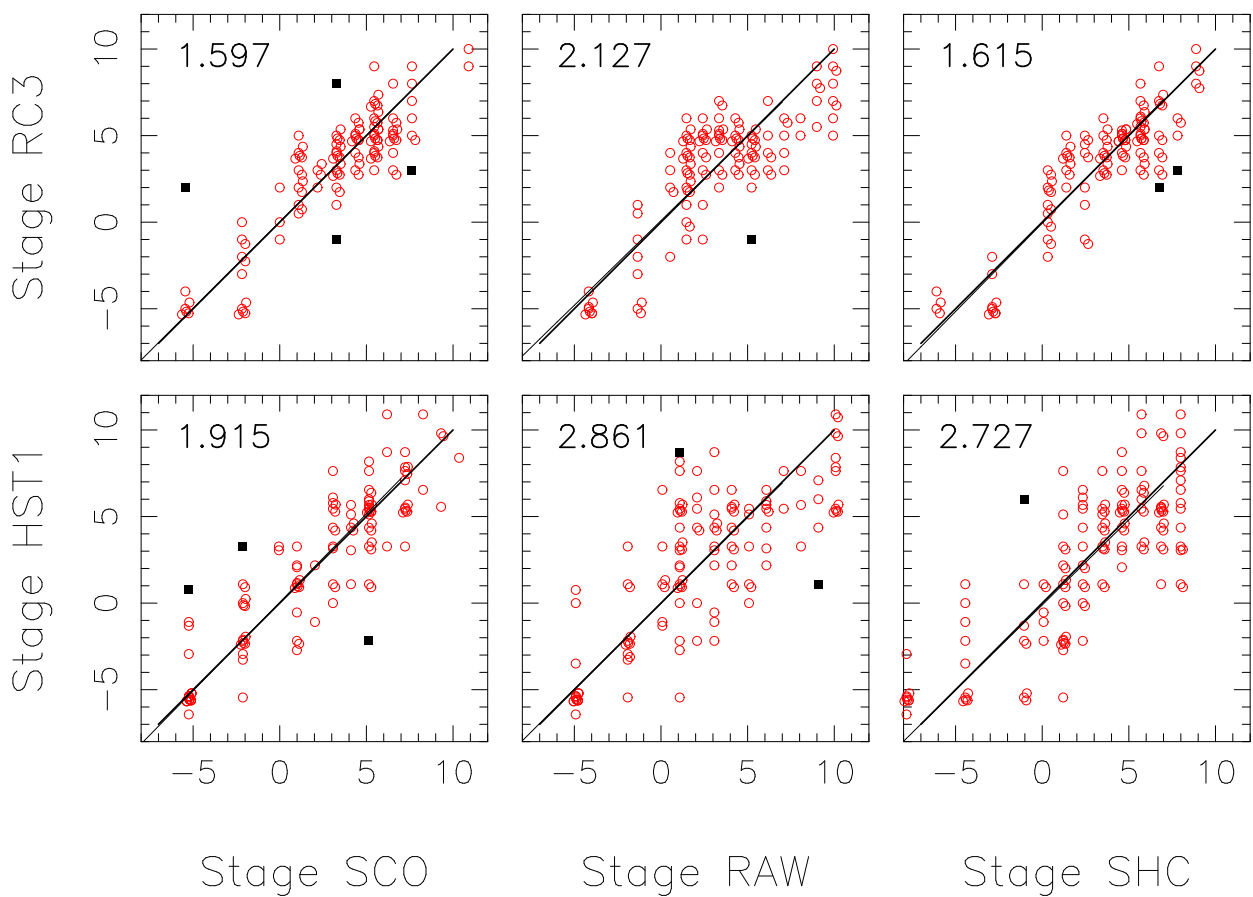}
\figcaption[f6.eps]{\sf \footnotesize Correlations between the revised Hubble 
         type stage estimates made by different classifiers. 
         Small scale and zeropoint 
         adjustments were made to each set in order to bring all 
         catalogs onto the system defined by RC3 stage estimates. 
         Using the method discussed in Odewahn and de Vaucouleurs 
         (1993) we derived the accidental error associated with 
         each classifier. \label{Tcompare}}
\end{figure}

 Our galaxy image libraries were independently inspected by 
three classifiers using an LMORPHO graphical galaxy classification 
tool.  The three classifiers were SCO, Windhorst (hereafter RAW), and 
Cohen (hereafter SHC), and each 
visually estimated the stage, family, and variety 
for each of the 272 galaxy images in the adopted sample. For 
a variety of reasons (insufficient \SN or resolution, high 
inclination, etc...) certain aspects of the classification 
(usually the family and variety) were uncertain and these 
were designated as unknown.  
Following the method used for the HST1 sample, we next 
inter-compared these type catalogs in order to establish the 
systematic and accidental errors associated with each set. 
The results of this are summarized in Table ~\ref{tab1} and 
Figure~\ref{Tcompare}.  For the local galaxy samples, we compared 
each classifier to the weighted mean types in the RHS contained in 
the RC3 and discussed in Buta \etal (1994). For the distant sample 
we compared each classifier's types with the weighted mean types of HST1.  
Although many of the HST1 types are indeed weighted means of several 
independent estimates, the majority are single estimates from SCO 
and hence we should regard the HST1 set as an independent 
(SCO-dominated) unit-weight catalog. 

  For both the local and distant galaxy samples, a preliminary 
linear regression analysis was performed to determine transformation 
relations for converting to the RC3 and HST1 systems respectively. 
An impartial regression analysis, using each catalog as the 
dependent variable, was used to determine the mean relationship 
between all possible catalog pairs.  For each sub-sample (local 
and distant), small scale and zeropoint shifts were 
applied to the SCO, RAW and SHC stage estimates to bring  
all classifications onto a uniform system. It is informative 
to note that the size and sense of the scale corrections 
were uniform by classifier irrespective of local or distant 
sample: SCO and RAW systematically classified galaxies 
later at the 8\% level, and SHC systematically classified
galaxies earlier at the 6\% level.  These scale adjustments were 
statistically significant at the $2\sigma$ level. The zeropoint 
shifts, statistically significant at the $3\sigma$ level, were 
general less than one step on the 16-step stage axis of the RHS  
(with the exception of a $-2.2$ step correction for the distant 
RAW sample).  In each case (local and distant) mean transformation 
equations were derived for the SCO, RAW, and SHC classification 
sets.  Following these small, but statistically justified, 
transformations a second correlation analysis was performed to
verify system uniformity and to compute the mean variance 
associated with each catalog comparison. This analysis is 
illustrated in the top panel of Figure~\ref{Tcompare} for the 
local sample and in the bottom panel for the distant (HST) 
sample. With all stage estimates transformed to a 
common mean system, we followed the methodology of 
Odewahn and de Vaucouleurs (1993) to compute the 
mean standard deviation of the $2\times2$ residuals in each 
comparison (these values are plotted in each panel of 
Figure~\ref{Tcompare}).
These estimates were combined via a system of linear 
equations to calculate the standard deviation associated with 
each individual catalog. The results, summarized in 
Table ~\ref{tab2}, show first that each classifier (SCO,RAW,SHC) 
is able to produce stage estimates for the local galaxies in 
the RC3-defined system with scatter of between 1 and 2 steps 
on the RHS. This is quite consistent with a 
similar study by Naim \etal (1995) using 831 galaxies typed from 
Schmidt images by 6 expert classifiers. As expected, the r.m.s. 
scatter estimated for the distant galaxy samples are somewhat 
larger, but still in the 1.5 to 2 step range. 

\begin{figure}
 \epsscale{1.0}
 \plotone{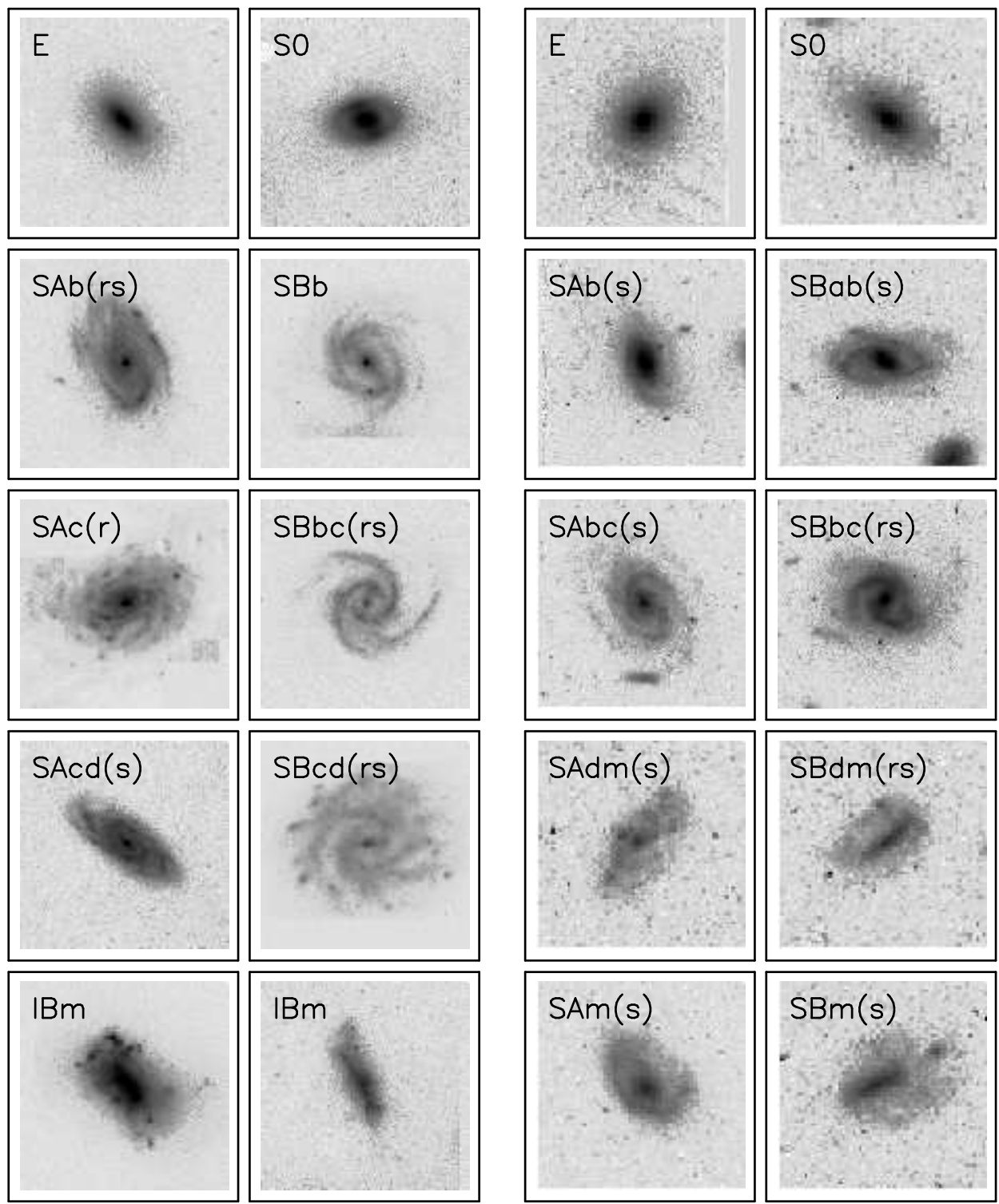}
\figcaption[f7.eps]{\sf \footnotesize Sample images for galaxies with weighted 
         mean revised Hubble types. The two columns on the left are taken 
         from a sample of local galaxies with types from the RC3, 
         and the two on the right are taken from a sample of distant 
         galaxies imaged with WFPC2 in F814W and types from SCO+RAW+SHC
         (see text). The nearby galaxy images have been processed to 
         yield \SN and resolutions measures to those comparably 
         found in HST images. The common morphological structures 
         described by the RHS are seen in both samples of galaxy 
         images. \label{Sample_gals}}
\end{figure}

 Using the systematic corrections and sample weights derived 
for the HST classification sets we have compiled a catalog 
of weighted mean revised Hubble types for our distant galaxy 
samples. In this way we produce not only a final catalog of 
higher quality types compared to that from any single classifier,
but we are able to derive error estimates for each galaxy. This 
last point is important from the point of view of assessing 
the quality of any automated morphological classifier. Galaxies 
whose morphological properties are uncertain to a number of 
human classifiers are most likely peculiar in some aspect and 
hence should not be included in the training or testing of a 
generic machine classifier.  Weighted mean types and errors for 
our local galaxy sample were adopted from the RC3. Representative 
galaxies from each of these samples, broken into stage and family 
groups, are shown in Figure~\ref{Sample_gals}.  

\subsection{A high weight set for family and variety classification}
  Although agreement between stage classifications was found to be 
quite satisfactory for developing adequate test/train samples for 
automatic stage classifiers, the uniformity of 
the family (barred vs. unbarred)
and the variety (ringed vs. non-ringed) classes was less than 
satisfactory. From the HST image samples, all 3 classifiers 
agreed on the family assignment only 30\% of the time and on the 
variety assignment only 25\% of the time. This is clearly due to
a lack of the  \SN and resolution required for such morphological 
classification in galaxy images. In reality, some of the ground-based 
images also lacked sufficient \SN to allow for the unambiguous 
assignment of family and variety assignment. Hence, for the purposes 
of developing a robust morphological classifier capable of dealing 
with these image properties we chose to develop a third, higher 
weight set of galaxy images. During the course of a large program to 
acquire photometric calibration fields for the Digital Palomar Sky 
Survey (Gal \etal 2000) one of us (SCO) initiated a program for 
non-photometric conditions of imaging large, bright well-classified 
galaxies from the list of "best-classified" galaxies in 
Buta \etal (1994). These galaxies had classifications agreed 
upon by several human experts and in fact represent proto-types 
of the various Hubble stages. Each galaxy was imaged in Gunn gri or 
Cousins B for 10 to 15 minutes with the Palomar 60-inch using 
CCD13. Conditions were generally not photometric, but thin 
cirrus and seeing no worse than 2.5 arcsec FWHM was tolerated for 
this program.  When possible, multiple images of 
the same galaxy were obtained so that we might characterize 
how accidental errors in the surface photometry will affect 
output from the automated morphological classifier. 
In Table ~\ref{P60table} we summarize 30 images in Gunn g and 
Cousins B of 20 different galaxies selected for experimenting 
with with automated family and variety classification.  



\section{PRACTICAL APPLICATION OF THE METHOD}
Here we describe the application of the method to a set of HST 
archival images. Using the results of Section 2,  we assembled 
samples of local and distant galaxies having weighted mean stage 
estimates and family classifications that at least two human 
classifiers agreed upon.  We show a representative sample of 
these galaxies in Figure~\ref{Sample_gals}. As we shall 
discuss below, the criteria for building a family classifier 
training set had to be somewhat liberal in order to compile 
a usable experimental sample.  For stage classification, a series of 
backpropagation networks using different input parameter 
vectors and network layer architectures were experimented
with. In the case of the family classifier, we used two 
different types ANN classifiers: a backpropagation-trained 
feed forward network, and a difference boosting neural 
network (DBNN) developed by Philip \etal (2000).  

\subsection{Stage Classification} 
 A set of 262 images were collected using the sample of galaxies
having $b/a > 0.4$ and weighted mean stage estimates from the 
analysis of Section 3. Most galaxy classification 
systems provide criteria for estimating the morphological 
type of an edge-on system, however because we are concerned 
here with the analysis of two-dimensional morphological 
structures, we chose to exclude such high-inclination systems. 
An LMORPHO task was used to perform automated surface photometry 
of each image, and then an interactive mosaic viewer was used to 
inspect the ellipse fit to the low surface brightness isophote 
in each galaxy postage stamp image. This was done to verify 
that no improper image processing errors occurred due to 
confusion from nearby images or other image defects. Ten 
galaxy stamps were rejected in this step, the majority of 
these being relatively faint HST-observed galaxies that 
extended too close to the edge of the WFPC2 field. The human 
classifiers dealt well with "masking" such edge effects when 
the visual classification of Section 2
was performed, however the sky fitting and surface brightness 
contouring routines of LMORPHO were unable to properly handle 
such situations. Future automated morphological surveys will of 
course have to deal more effectively with this type of error, but 
for the purpose of this work we chose to simply delete these 
galaxies from any machine classification experiments.  
 
 For each of the 252 postage stamp images of well-classified 
galaxies remaining in our experimental sample the full set of 
morphologically-dependent Fourier 
image parameters discussed in Section 3 were computed 
with LMORPHO. A series of backpropagation neural networks 
were trained following the procedures outlined in Odewahn (1997) 
using 3 different types of input vector sets and 6 different 
network architectures. In every training case, two hidden 
layers of the same size were used to map n-dimension input vectors 
to an 8-node output layer, one node for each 2-step interval 
of the stage axis of the RHS.
In practice, a final type was assigned for each input vector 
using the weighted mean value of the output node value. This 
allows one to assign not only the highest weight classification, but 
the rms scatter in this node-weighted mean value can be used 
to assign a classification confidence. Hence, while any output 
pattern will result in the assignment of an estimated type, 
patterns where most of the output signal is carried in one major 
node will have a high confidence assignment, and patterns where 
the output pattern is distributed over many nodes will be 
assigned low confidence. 

\begin{figure}
 \epsscale{1.0}
 \plotone{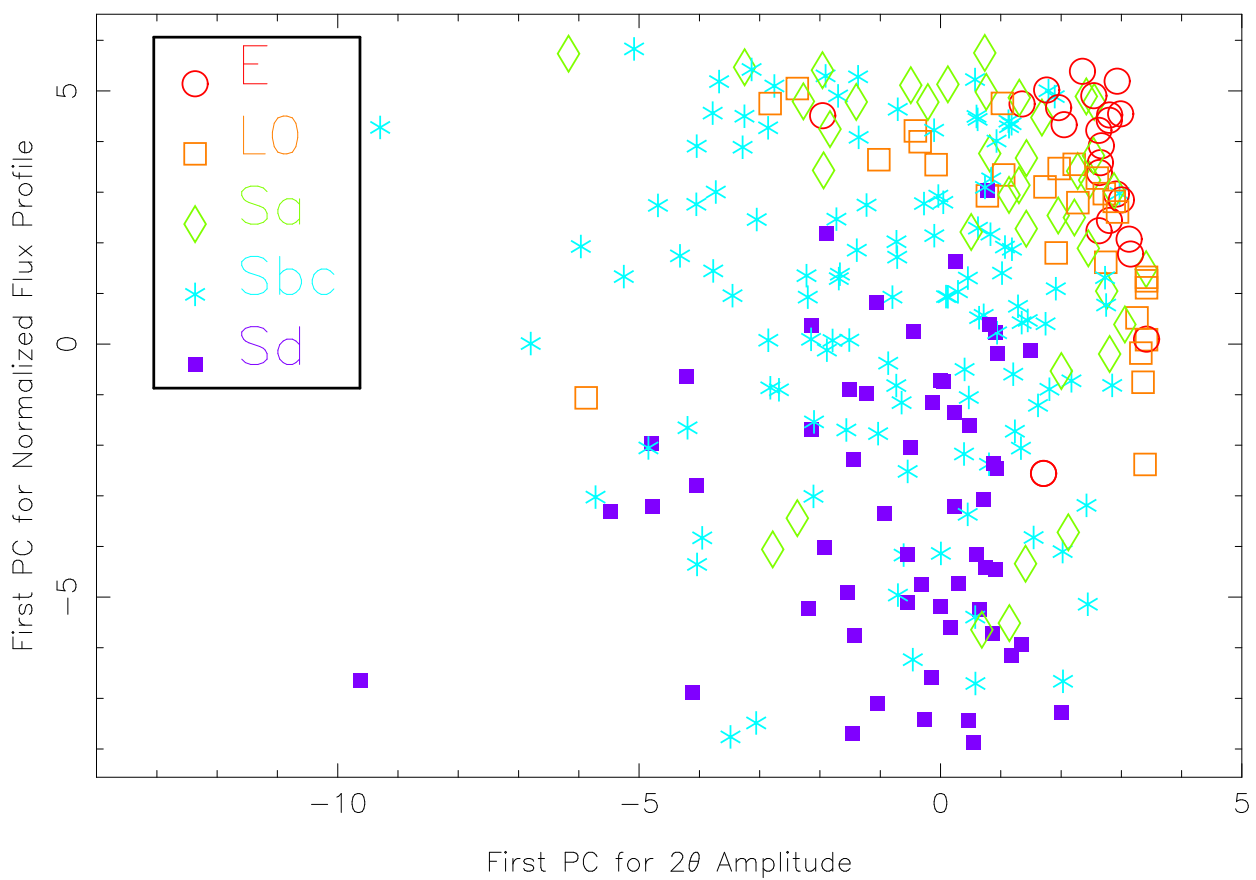}
\figcaption[f8.eps]{\sf \footnotesize The first principal components 
         formed from an analysis of the $2\theta$ amplitude and 
         normalized flux profiles from 246 local and distant galaxies. 
         We see a clear segregation by revised Hubble 
         stage (normally referred to as the Hubble type). A variety of 
         independent and similarly fuzzy Fourier-based image parameter 
         may be combined via the ANN classifiers described in the text 
         to predict stage values. 
         \label{pars_STAGE}}
\end{figure}

\begin{figure}
 \epsscale{1.0}
 \plotone{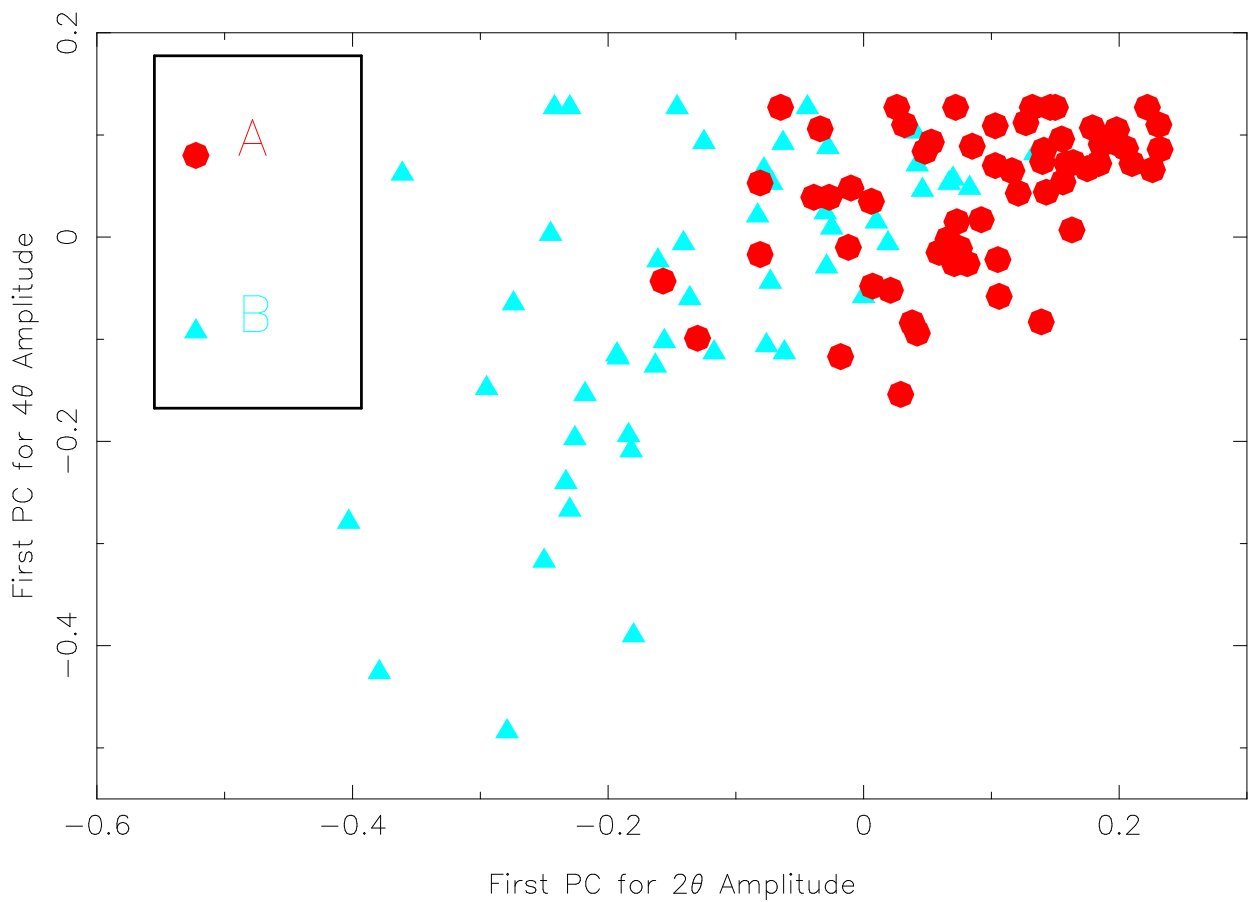}
\figcaption[f9.eps]{\sf \footnotesize The first principal components formed 
         from an analysis of the $2\theta$ and $4\theta$ profiles from 205 local 
         and distant galaxies. We see a fuzzy segregation by revised Hubble 
         family (barred vs. unbarred). The unbarred (A) galaxies 
         preferentially occupy the upper region of this space. The AB 
         galaxies cover an intermediate region of this parameter space, 
         and for clarity we plot only the A and B galaxies. 
         \label{pars_FAMILY}}
\end{figure}

  The image parameter catalogs were collected into a single 
binary catalog which can be used in LMORPHO to interactively 
generate symbol-coded parameter space plots like those shown 
in Figures ~\ref{pars_STAGE} and ~\ref{pars_FAMILY}.  
With this package the user can quickly view many different 
parameter spaces and judge which one shows the largest 
degree of type separation (whether stage or family) by 
determining how well the different type-coding symbols are 
separated in the plot. An important feature in this graphical 
tool is that any point may be marked and used for a variety 
of uses. In one mode the user can choose to view on a real-time 
basis the galaxy image of any selected data point. In this way
one is able to determine if a particular outlier is due to 
improper image processing or some truly unique morphological 
circumstance. In the latter case, such data are retained for 
use in classifier training. In the former, these patterns would 
be rejected from use in gathering training and testing samples. 
As the images used in this particular series of classification
experiments were generally of high quality, fewer than 5\% of 
the original 252 postage stamps were rejected in this manner.  
For only a few hundred galaxy images (or even less than a 
few thousand) this process of parameter space inspection is 
trivial to complete in a few minutes. Finally, we made a 
preliminary set of ANN training runs using different input 
parameter sets as well as subdividing the training/testing 
data in different subsets. Through this process we uncovered 
an additional 28 objects that consistantly gave highly discrepant 
training results, \ie the image parameter combinations for these 
sources were highly abnormal compared to the bulk of the data, and 
these sources were rejected from further experiments. This represented 
about 9\% of our original training set, but determining the source of 
peculiarity for these galaxies will require a larger, more diverse 
collection of images for future experiments. 
We determined a number of useful input parameter combinations that 
showed good segregation by morphological type and a series of 
different image parameter set patterns were formulated. Each 
type of pattern would form the input layer to an ANN classifier. 
For clarity, we assigned a running integer value to each type 
of input, and we refer to this as the feature set number. We indicate
the feature set number for each training exercise in column 8 of Table 6. 
  
In Table~\ref{ParList} we summarize the LMORPHO image parameters
that were selected for the stage and family classifiers developed 
in this paper. Since different image parameters are used in each 
classifier, we list in column 2 the feature set number for each 
input pattern using each image parameter.  

The selection of training and testing samples was carried out in 
the same manner for all input vector types. As summarized 
in Table~\ref{Tbreakdown} all galaxies were binned into the 
8 type bins corresponding to the 8 nodes of the ANN output 
layer. Each such bin was divided into two sets: one for training 
and one for testing. One is always tempted to use more galaxies 
in the training sample so that the network has a larger 
number of patterns for generalizing the problem. One danger 
in backpropagation training is over-training: beyond 
some number of training iterations the network weights could  
be adjusted so as to simply memorize the training pattern 
sets as opposed to generalizing the mapping of input vectors 
to output classes. To prevent this, one 
should use an independent test data set to judge classifier 
performance. The backpropagation code in LMORPHO computes 
classifier statistics for both a training sample and a testing  
sample at each user-specified iteration in which the updated 
ANN weights are stored. In a post-training phase, each set 
of statistics is inspected by the code to determine the optimal 
training cycle. The algorithm used to isolate this iteration 
uses the slope and rms scatter of the correlation between the 
target stage prediction and the ANN stage prediction. Both the 
training and testing data sets are used. The case of over-training
is detected by a progressive drop in the rms scatter for the 
training data with a flat or even increasing rms scatter for 
the test sample. An optimal training cycle is selected that 
minimizes the rms scatter for training data before 
over-training occurs. Additionally, the slope of the 
linear regression must be close to unity within some user-specified 
tolerance, usually set at 20\%. 

We summarize the results of our backpropagation training to 
develop Fourier-based ANN classifiers in Table ~\ref{StageClassStats}.
The slope ($\alpha$) and scatter ($\sigma_{1}$) about a linear regression 
between the ANN-predicted and the targeted mean Hubble stage are tabulated 
for both the training and testing data sets for the selected optimal training 
epoch. Additionally, as with the type comparison analysis in Section 3, 
we compute a scatter ($\sigma_{2}$) based on direct differences 
between the ANN and target values following Odewahn and de Vaucouleurs (1993).
This statistic reflects more honestly the scatter to be expected for the 
user of such an ANN-derived catalog.  As expected, scatter among the training 
data is systematically lower than for the test data, since the error 
function being minimized in the backpropagation training is formed with 
the training data. However, the scatter derived for the test data is 
usually about 2-steps on the RHS stage axis, and hence very reasonable 
for most scientific pursuits with morphological data. 

\begin{figure}
 \epsscale{1.0}
 \plottwo{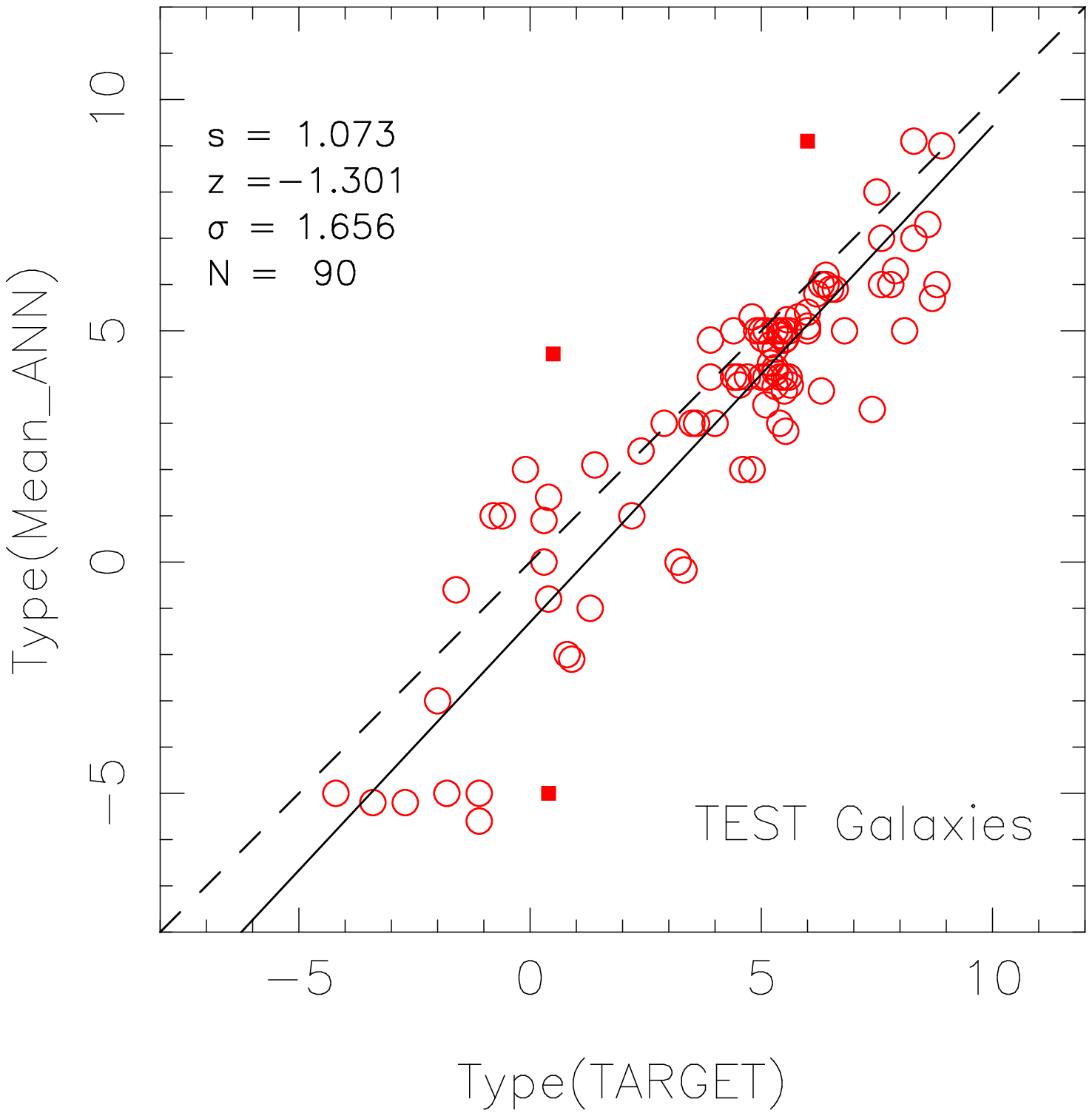}{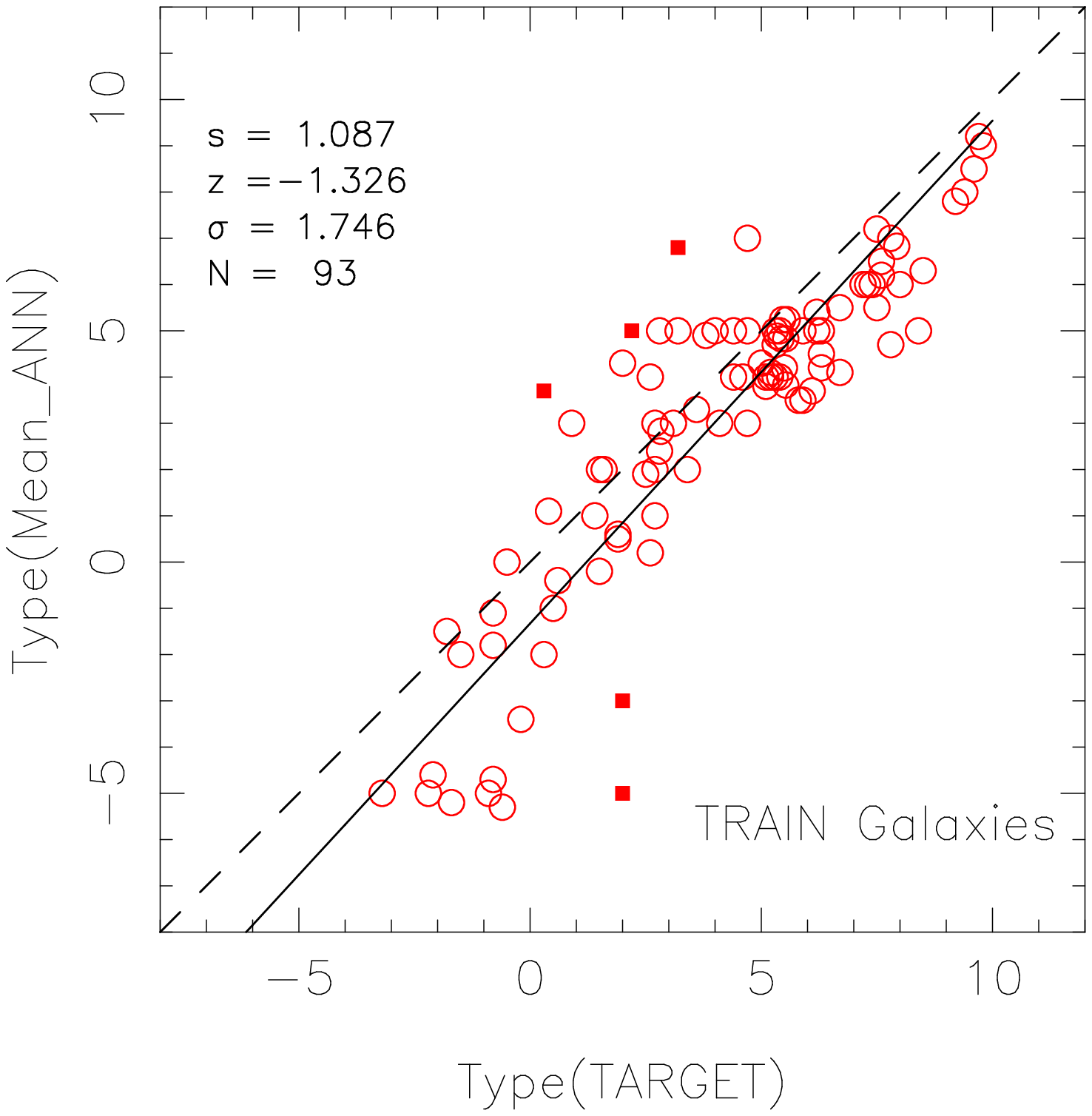}
\figcaption[f10a.eps,f10b.eps]{\sf \footnotesize (a) [Left] The mean 
          stage estimates from 4 independently trained 
          ANN classifiers as a function of target type as given by 
          the catalog of weighted mean visual stage estimates. As discussed 
          in the text, each ANN classifier is fed a combination of
          principal components based on an eigenvector analysis of 
          Fourier amplitude profiles. The linear regression indicated 
          by the solid line departs systematically from the dashed 
          unity line but gives a reasonable y-axis residual scatter of 1.6 
          steps on the 16-step revised Hubble stage axis. [Right] The same 
          correlation with the exception that we plot only data from galaxies 
          used as input patterns during backpropagation training to establish 
          ANN weight values. In both cases, the the small box symbols are
          data rejected after two cycles of 3-sigma rejection in the
          linear regression fit.
          \label{net_results}}
\end{figure}

In Figure~\ref{net_results} we compare mean neural network classifier types 
from 3 different ANNs (marked with asterisks in Table ~\ref{StageClassStats}) 
to weighted mean visual types for two sets of data: the 
training data and the testing data. To summarize, the test data, comprised of sets 
of input classification parameters and their corresponding target 
types, were never presented to the ANN classifiers during backpropagation 
training to establish the network weight values. As such, this data 
sample represents a fair comparison by which we can judge true 
network performance. We include in Figure~\ref{net_results} the same 
correlation for the training sample. In summary, the Fourier-based 
stage classifiers developed in this experiment clearly perform as 
well as a human classifier. These results are certainly encouraging 
if our goal is simply to estimate revised Hubble types for a 
large number of digital images, but a more important point must 
be stressed. We have shown that our method of parameterizing a 
galaxy image preserves the information content needed to emulate 
the process used by a human classifier. As with the bar classification 
work described below, our ultimate goal will be to use such a 
digital image analysis to search for more direct correlations 
among galaxy properties and ultimately develop a more physically 
meaningful classification system beyond the RHS.  

\subsection{Family Classification} 
As discussed in Section 3, the visual classifiers divided the 
family estimates into the three bins of the RHS: A, AB, B.  
The difference between each class is somewhat vague and subject 
to personal bias. It is not surprising that we found very few 
cases where all three classifier agree that a galaxy was of 
the AB family class. As was discussed for Figure~\ref{MFPROF1},
the Fourier profiles for AB galaxies are, in the mean, intermediate 
between the A and B galaxies. However, the AB galaxies show a 
markedly large variance which is probably due to the wide 
variance in AB classification criteria used by the human 
classifiers. The presence of such uncertainty made it impossible 
to train an effective 3-division stage classifier. For the 
present work, we approached the problem using an unambiguous
set of A and B galaxies and attempted to develop a robust 
classifier for identifying them using the Fourier-based 
eigenvector approach discussed in Section 3. 
When larger samples of morphologically classifiable galaxy 
images are collected, it is anticipated that a more thorough 
study of bar strength parameters will allow the development 
of a more sophisticated automated family classifier. 

For the present set of classification experiments we gathered 
images of 71 A galaxies and 61 B galaxies drawn from the analysis 
of Section 3. In this case, any galaxy which had family classifications
that were agreed upon by at least 2 of the 3 human classifiers were 
adopted for use. To clarify the problem, we used only A and B classes 
for classifier experiments. After a series of parameter space 
inspections, such as those described in the stage classification 
section, we chose to use the first principal components of the 
$2\theta$ and $4\theta$ amplitude profiles. We found that additional 
discrimination was added in many cases using the first principal 
component of the $4\theta$ phase profile and adopted this for 
use in the classifier input vector. It was found that including  
parameters that are highly correlated with bulge to disk ratio 
(\ie the principal components of the normalized flux profile) did
not add significantly to detecting the presence of a bar. This was 
surprising in that it is well known that bar shapes and properties 
are correlated with Hubble stage, and hence bulge to disk ratio
(Elmegreen and Elmegreen 1986). In future work with much larger 
samples of galaxies, the use of terms related to the shape 
of the surface brightness profile may prove useful for this 
reason: pattern classifiers can better interpret the meaning 
of morphological shape parameters like the first principal 
components of the $2\theta$ and $4\theta$ amplitude profiles  
if some information related to Hubble stage is provided. For 
the present small sample of potentially useful training samples, 
we decided to exclude the use of surface brightness profile shape 
parameters for the training of a family classifier.  

  We added two simple global image parameters that contribute 
information about bar presence that were independent of the 
Fourier model approach. In determining 
the shape and orientation of the optimal elliptical aperture for 
measuring each galaxy, LMORPHO computes a series of ellipse fits 
to progressively fainter surface brightness levels. The initial 
surface brightness levels are high and generally sample the bright 
inner region of a galaxy where the bars are found.  It was 
found that simple parameters using these elliptical 
isophote parameters could be formed that gave useful information 
about the presence of a bar that would be independent of the 
image models derived with the Fourier method. We define
the minimum axis ratio of the series of ellipse fits to be
$B_{P1}$, and the axis ratio of the largest ellipse (i.e. the 
fit to the lowest surface brightness level) to be $B_{P2}$. 
In general, a barred galaxy will have $B_{P1} < B_{P2}$, but 
with the condition that the major axis length of the 
isophote measured  for $B_{P1}$ is significantly smaller 
than that measured at the isophote for $B_{P2}$. To discriminate 
this condition we formed the parameter $B_{P3} = r_{1}/r_{2}$, 
where $r_{1}$ is the semi-major axis length measured at $B_{P1}$
and $r_{2}$ is the semi-major axis length measured at $B_{P2}$.
Additionally, we formed $B_{P4} = B_{P2}/B_{P1}$, to measure the 
contrast of the ellipticity variation in the galaxy image. It was 
found that for many galaxies, a plot with $B_{P4}$ on the y-axis 
and $B_{P3}$ on the x-axis places B galaxies in the upper left
corner and A galaxies in the lower right corner. Of course, there 
is a rather substantial overlapping zone in the 
low $B_{P4}$ and low $B_{P3}$ region of this plot. Hence, while 
such a simple scheme can not discriminate all A,B galaxies, it 
can provide high weight information for some systems. Simple 
initial ANN classification experiments provided evidence that 
these parameters, when added to the Fourier-based parameters, 
produced better discrimination for A,B samples and hence we 
included them in the training of our final family classifier. 

 Two types of family classifiers were trained with the input 
vector parameters shown in Table~\ref{ParList}. First, the 
backpropagation network code used the previously discussed 
stage classifiers was used to train a network consisting of 
2 hidden layers having 12 nodes each. This network mapped 
an 8-element input vector to a 2-node output layer. Patterns 
with more power produced in output node 1 were classified as 
A, and patterns with more power produced in output node 2 were 
classified as B. Many of the dimensions in this 8-dimensional 
classification space produce a good split between the A and 
B populations and hence the backpropagation training converged 
relatively quickly. As with the stage classifiers, a sample 
of pure test patterns was retained and used to assess network 
performance at each stage in the backpropagation weight 
update process. This procedure gives us a truly independent 
check on the classifier performance and an effective means 
of guarding against over-training. A total of 30 training 
epochs were used in the backpropagation training, and 
epoch 15 was selected as providing optimal performance with 
a success rate performance of 92\% among training patterns 
and 85\% among test patterns. In a second experiment the 
DBNN method of Philip \etal (2000) was applied to the identical 
training and testing patterns as for the backpropagation ANN.
After boosting, the training set was found to 
produce a 94.1\% successful classification rate and the 
test set was found to produce a success rate of 87.5\%. The 
performance of the DBNN classifier was found to be marginally 
better than the backpropagation-trained network, however a 
large sample of training data will be needed to assess if 
this improvement is significant. Nevertheless, both methods
produced automated family classifiers that are able to 
discriminate bar presence with a roughly 90\% probability of 
success.     

\subsection{Systematic Effects with \SN and Resolution}
  Image quality plays a crucial role in determining the extent to 
which morphological classification can be performed on a galaxy. 
A gradual decrease in image resolution and \SN will systematically 
degrade the quality of morphological classifications, both by 
human and machine methods. With stage estimation, one might expect 
that we lose the ability to differentiate the bulge and 
disk components, making it extremely difficult to work at 
the early stage of the Hubble sequence (\ie differentiate 
E, S0 and early spiral systems). At the late Hubble stages, 
we lose many of the low surface brightness features in the 
out disk regions that are crucial for differentiating 
among the Sd, Sm and Im systems. With family classification, one 
can expect that gradual image degradation will make it 
impossible to detect the presence of a bar, much less characterize 
the properties of the bar (see van den Bergh 2001). In the 
past, human classifiers, in particular those dealing with 
low resolution and low \SN Schmidt plate images (Nilson 
1973, Corwin \etal 1985), have generally taken such effects 
into account by dropping the amount of detail in the literal 
type assigned to a galaxy. In other words, a system might be  
classified simply as "S" rather that "SBc", if the image 
quality is sufficient only to differentiate the difference 
between early-, mid-, or late-type galaxy. Such effects 
must be accounted for in any automated system geared towards 
the recognition of morphological features. One must know if there 
is a strong systematic bias towards, for instance, missing 
bar or spiral features as resolution decreases. Perhaps if 
such effects are well understood we may hope to correct 
large statistical samples in some systematic way. At the 
very least, we must quantify the levels of low \SN or 
resolution that can be tolerated before the quality of  
automated classifications falls below some minimum 
tolerance required by a given science goal. 

We chose to define resolution as the number of pixels  
in a galaxy image contained within the isophotal ellipse used to 
integrate the isophotal magnitude in LMORPHO. This is a reasonable 
approach for HST images where, for most filters, each pixel approaches  
the resolution of the optical image. For ground based images we 
should use the size of the seeing disk, which is generally several 
times larger than the pixel size. In our present study all of the 
ground based images have been block averaged to produce pixel 
sizes that are comparable to the seeing disk and hence we can use 
this uniform resolution definition for all of the galaxies 
analyzed. As for defining the \SN in a galaxy image we used two 
approaches. In the first, we simply compute the ratio
of the  mean signal (above sky) per pixel to the standard 
deviation of the local sky measure. In the second approach, we 
define the mean signal, $S$, to be the zeroeth order term 
in the Fourier series fitted to the azimuthal distribution 
in each model annulus, and the noise, $N$, is the r.m.s. 
scatter about that fit. This latter approach incorporates 
the Poisson noise associated with the detected galaxy signal 
and accounts for large scale structural changes across the 
azimuthal profile. In practice, the two methods had very 
different ranges, but were found to be well correlated. 
The method 1 approach yielded \SN estimates in the 
range 100 to 1000, and the Fourier-based method produced 
values in the range of 3 to 20. It was found that the early-type 
galaxies, having little large scale structure in their
images, produced the tightest correlation between these two 
types of \SN estimators. For practical reasons, we used 
the more easily computed method 1 values in the image classification 
experiments described in this section. 

To characterize how decreased \SN and resolution will affect the 
quality of our Fourier-based ANN classifiers, we selected sets 
of galaxies having extremely high \SN and resolution. These systems 
were found to be classified correctly by the stage and family classifiers 
discussed in the previous sections. Most of these images 
were taken from the ground-based datasets obtained 
with the Palomar 60$''$ discussed in Section 3, but a few of the 
highest resolution HST images were also included. For samples 
having well classified stage and family types, we selected 
galaxies with images that had at least 2000 pixels contained 
within the optimal elliptical aperture. These sets of images 
were then systematically re-sampled using an LMORPHO package 
designed for this experiment to produce postage stamp images 
of galaxies having progressively lower \SN and resolution. 
As this exercise was designed to study the general systematic 
effects of image degradation on morphological classification, 
no attempt was made to model the noise properties of 
a given detector, like WFPC2 or STIS (as in Odewahn \etal 1997). 
For each re-sampled image, an extra component of gaussian 
noise was added to the original image. We modeled decreasing 
resolution in two ways. First, we consider the case of high 
sample rate with image blurring, the case typically encountered 
with ground-based observations of nearby galaxies. In the second, 
and more relevant case, we considered a low sample rate with 
high optical resolution, \ie the image suffers from a 
high degree of pixelation. To simulate this effect, we simply 
block averaged the original galaxy images to a lower image 
sampling rate. This latter effect is well known to 
anyone who has classified large numbers of distant galaxies 
on WFPC2 images and is the one, as we will show, which dominates
our ability to determine distant galaxy morphological classification. 

\begin{figure}
 \epsscale{1.0}
 \plottwo{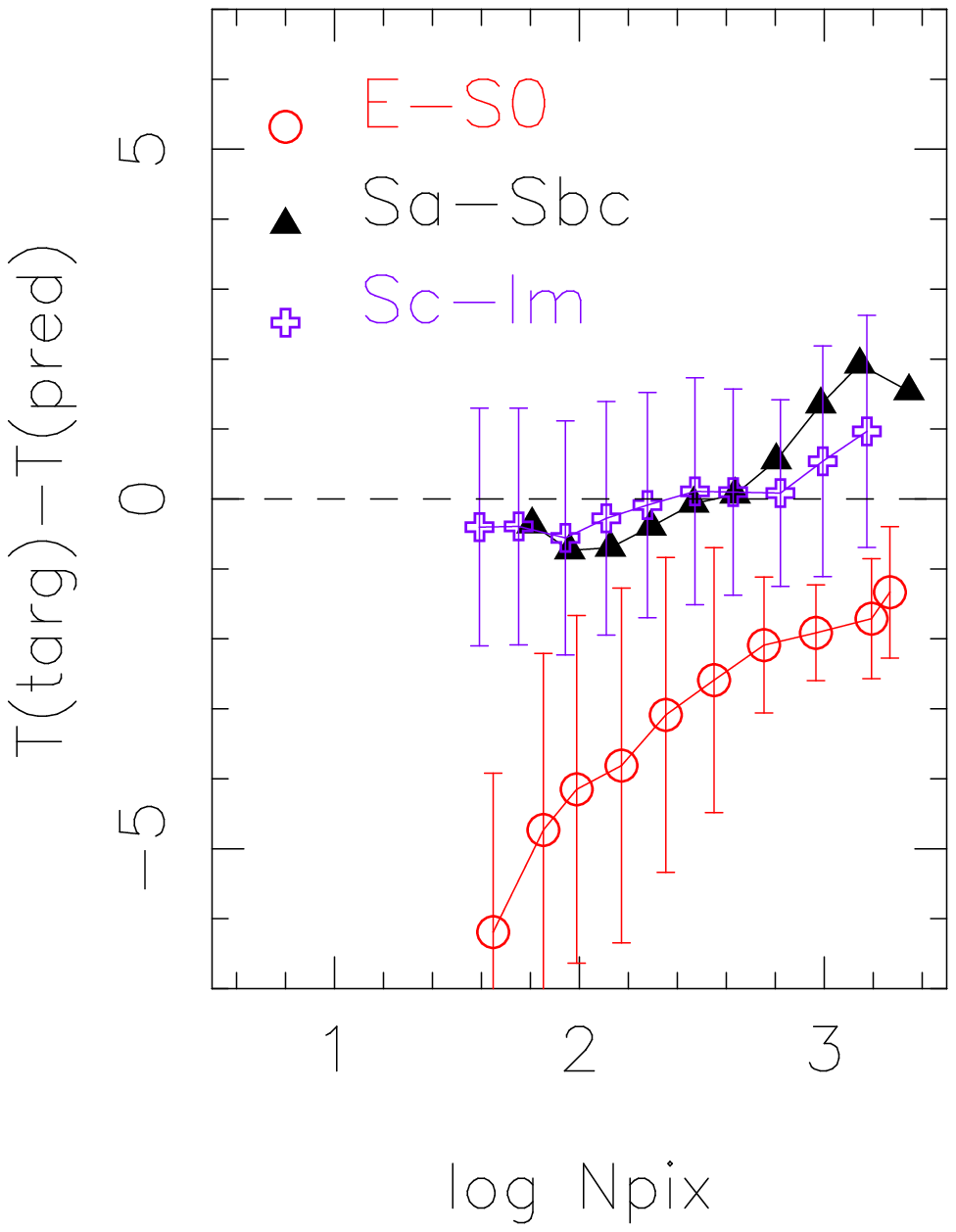}{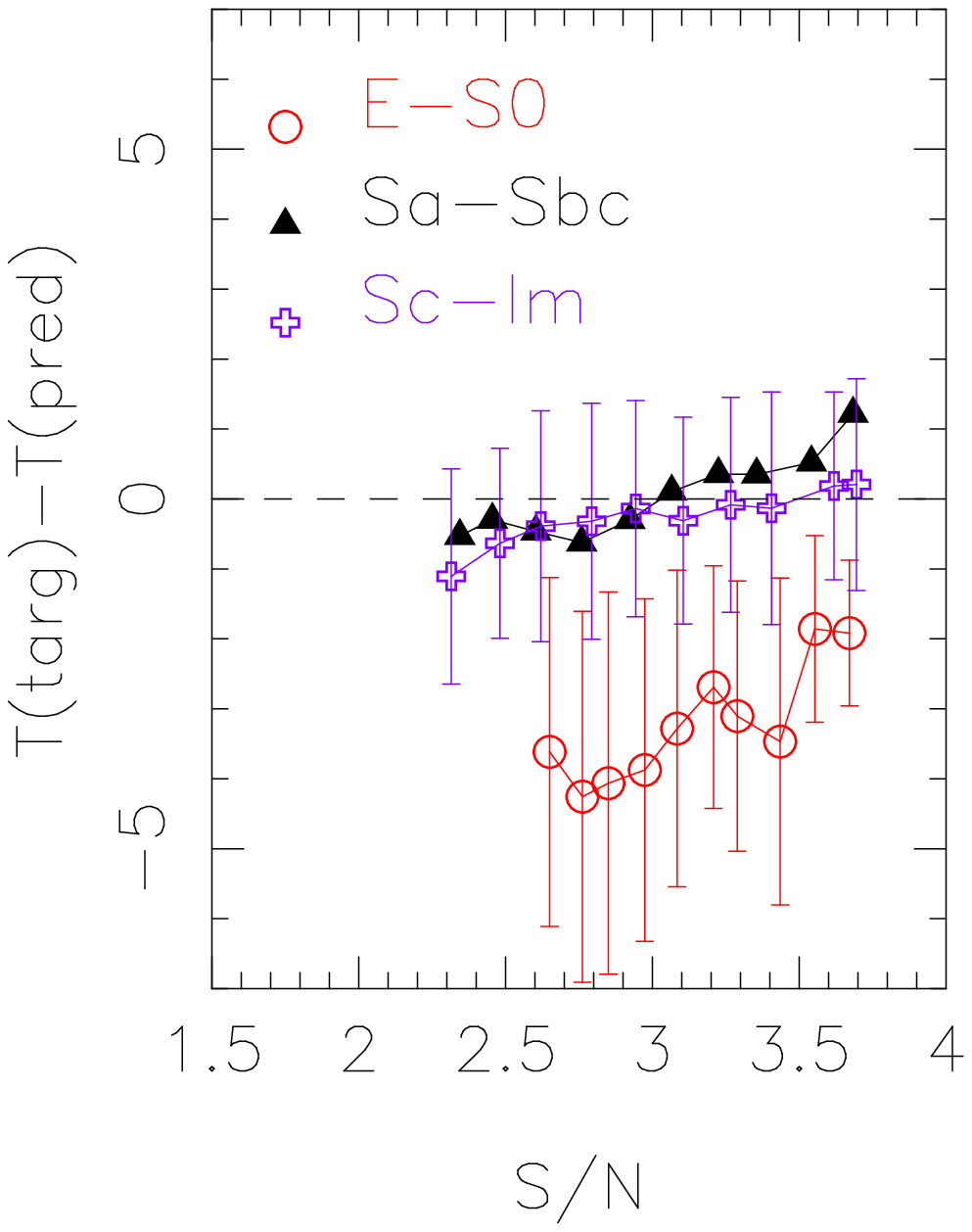}
\figcaption[f11a.eps,f11b.eps]{\sf \footnotesize Resolution and \SN in a set 
         of 261 galaxy images have 
         been systematically degraded using repixelation and 
         the addition of a gaussian noise component.
         The resulting 1901 images were processed with the ANN classifiers
         discussed in Section 4.2 to estimate Hubble stage. We 
         plot the mean trend in stage residual (target-predicted) for 
         three intervals of target galaxy stage as a function 
         of resolution as measure by the number of image 
         pixels (left) and \SN (right). In each case we 
         have used overlapping x-axis bins of width 0.2 dex to 
         demonstrate the smooth degradation in classifier 
         performance as image quality is lowered. Spiral and irregular 
         stages show a moderate trend in positive mean offset (ANN 
         classifies earlier) and increased rms scatter with decreasing 
         resolution and \SN. The E+S0 systems, as one expects from 
         viewing Figure~\ref{net_results}, exhibit a mean negative 
         offset (ANN classifies later) of 2-3 Hubble steps at even 
         the best image quality for the present samples. Additionally, 
         this mean offset and increasing rms scatter in T-types is  
         clearly steeper than for spiral and irregular galaxies.
         \label{systematics_stage}}
\end{figure}

The results for the backpropagation ANN stage classifiers are 
summarized in Figure~\ref{systematics_stage}. For this experiment 
a simple success rate calculation was not appropriate. We desire 
to know not only whether a classifier fails to predict the 
correct stage estimate within some number of type bins, but 
in which direction a misclassification occurs. If the classifier 
consistently pushed stage estimates in an early (towards E,S0) 
or a late (toward Im) direction then a substantial bias will be 
introduced in the scientific interpretation of the ANN-generated 
morphological catalog. Hence, we chose to compute the trend 
in type stage residuals, computed in the sense of $T(target)-T(ANN)$, 
as a function of resolution and \SN, where $T(target)$ is 
the true galaxy type. These trends were computed using 
overlapping bins in resolution or \SN, with the mean stage residual 
and the rms about that mean computed in each bin. A set of  
1901 re-sampled images were processed with the ANN classifiers
discussed in Section 4.2 to estimate Hubble stage.  In 
Figure~\ref{systematics_stage} we show trends in stage residuals 
for three sets of target stage intervals: E-S0, Sa-Sbc, and Sc-Im. 
In each case we have used overlapping x-axis bins of width 0.2 dex, and 
we expect to observe the degradation in classifier performance as 
image quality is lowered. Spiral and irregular stages show a moderate 
trend in positive mean offset (\ie the ANN classifies these galaxies 
with an earlier value) and increased rms scatter about the mean
with decreasing resolution and \SN. Of more significance, E+S0 galaxies  
exhibit a mean negative offset (\ie the ANN classifies these systems 
to be later than the target value) of 2-3 Hubble steps. This occurs 
for even the highest resolution and highest \SN galaxy images in our 
present sample.  Additionally, this mean offset and increasing rms 
scatter in T-types for the E+S0 galaxies is clearly steeper than for 
spiral and irregular galaxies.  A variety of network 
architectures and input parameter vectors were 
experimented with in order to correct this problem. Little improvement 
was obtained due in large part to the lack of a sufficiently large 
sample of images for such galaxies. As discussed in Cohen \etal (2001), 
E+S0 in the field are relatively rare for \Iband \cle 22, and 
hence a very large number of HST parallel observations are needed 
to build up a significant image library.  Work is currently underway 
to collect such images from the many archival HST observations 
of moderate-redshift clusters. Clusters contain not only a large 
number of galaxies in a single WFPC2 field, but the morphological 
fractions of these galaxies are skewed to the early-types. The analysis 
of such large samples will potentially eliminate the mean negative 
stage offsets observed for the E+S0 samples of  Figure~\ref{systematics_stage},
however it is doubtful that we can hope for a comparable decrease in 
the rms of such stage classifications in the low \SN and resolution 
regimes. It is clear that images of sufficiently high quality 
are needed to differentiate many of the subtle structural 
properties of early-type galaxies, as is clear from the 
work of Im \etal (2000). The larger field size, higher quantum efficiency, 
and especially increased resolution of ACS, compared to WFPC2, will 
make this the instrument of choice on HST for identifying large samples 
of moderate-redshift E+S0 systems in the near future. 
 
\begin{figure}
 \epsscale{1.0}
 \plotone{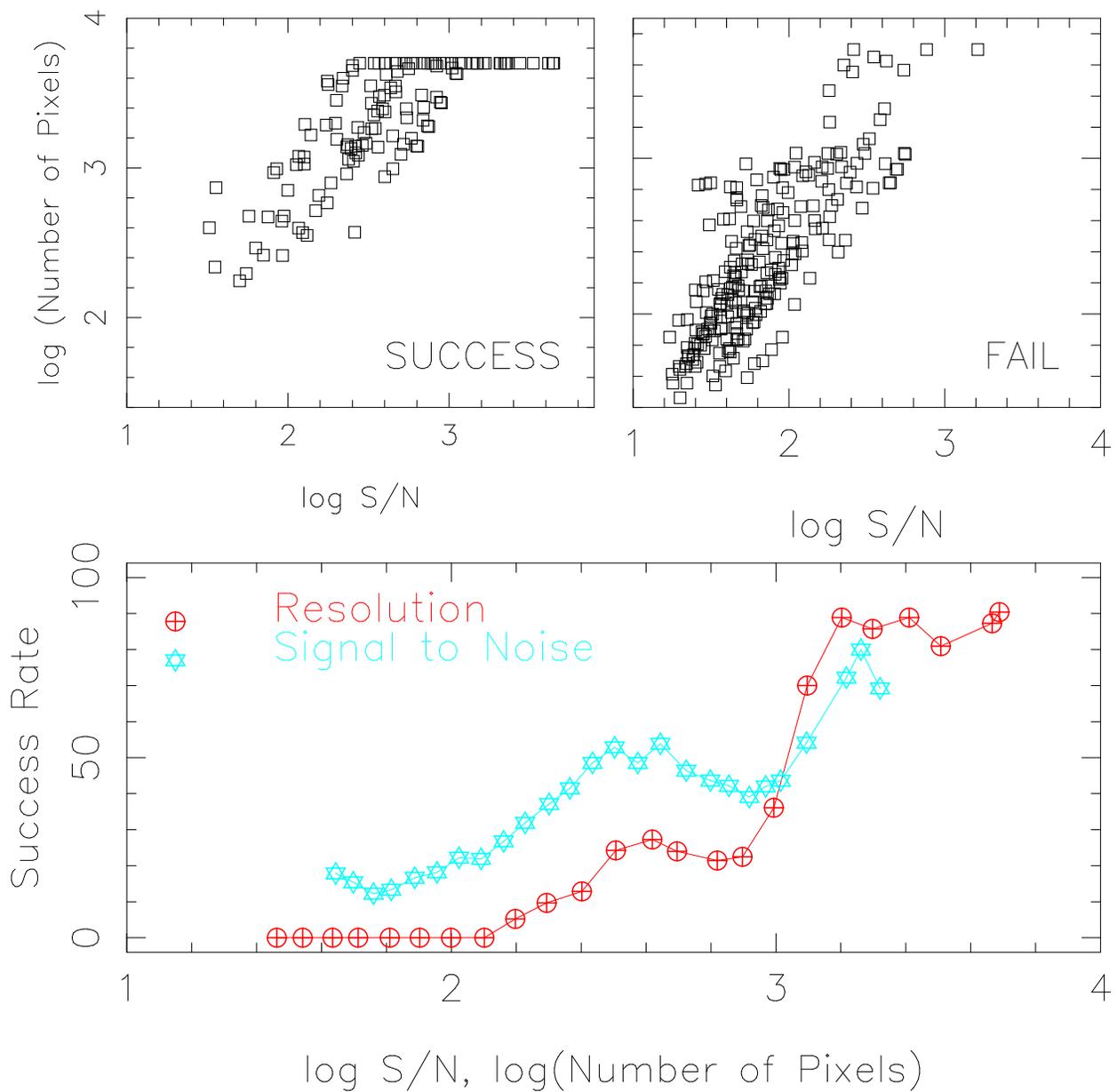}
\figcaption[f12.eps]{\sf \footnotesize The resolution and \SN of a set 
         of excellent quality 
         images of clearly barred galaxies (family=B) have been 
         systematically degraded to produce trends in \SN 
         vs. resolution (as measured by number of pixels) for 
         the cases of classifier success (left) and failure (right). 
         The resulting mean trends in classifier success with 
         \SN (star) and resolution (cross) are shown in the lower 
         panel.  \label{systematics_bar}}
\end{figure}

The results for the backpropagation ANN family classifier are 
summarized in Figure~\ref{systematics_bar}. Unlike the stage classifiers,
our family classifiers predict only two states: A or B. Hence, we 
are able to characterize the output results as success or 
failure cases. We therefore chose to plot the distribution of 
\SN and resolution for ANN-predicted classes for a sample of 
barred galaxies that agree with the target value and for those 
that fail to agree. As expected, in Figure~\ref{systematics_bar} the  
success cases cluster in the high \SN, high resolution area of the 
plot, and the failure cases cluster most heavily in the low \SN and 
resolution region. For clarity, we also plot the success rate trends
binned by \SN and resolution.  
For identifying the presence of a bar, it is quite 
clear from these mean relations that some critical image resolution 
is required. For our present samples and Fourier-based ANN 
classifiers, we are unable to effectively identify barred systems 
at better than a 70\% success rate until we obtain galaxy images 
with at least 1000 resolution elements ($\log N_{pixels}$ \cge 3, where
$\log N_{pixels}$ is the number of image pixels above the isophotal 
threshold). The trend with \SN is less steep, but a mean \SN (per pixel) 
of around 1000 is also required to reliably identify bars. 
The same experiment was 
carried out for a sample of unbarred (A) galaxies with different,
but expected, results. In this case, galaxies that are successfully 
classified as A at high \SN and resolution generally retain 
that classification at low \SN and resolution. In other words, 
the ANN classifiers rarely turn A galaxies into B galaxies, as is 
true for human classifiers. In general, as we lose image resolution 
we begin to loose the ability to identify morphological bars, but 
we do not tend to contaminate B galaxy samples with misclassified 
A galaxies, an important point when considering the the frequency 
of barred systems at high redshift.  

\section{SUMMARY}
  In this work we have developed a methodology for classifying galaxies 
based on the strength of Fourier components in the azimuthal light 
distribution. We have used pattern recognition techniques to develop 
machine classifiers which map the Fourier mode information to the 
revised Hubble system, a well known galaxy classification system. We 
must stress that the basic motivation here is not simply to create a 
machine-based replication of the revised Hubble system. The 
RHS was developed to describe a variety of 
properties, some of which are directly related with two-dimensional 
morphological structures in galaxies. We therefore have chosen to 
use this classification system to guide our development of 
new quantitative classification systems. We have 
demonstrated here a method of extracting galaxy image information 
in a way which permits us to automate the recognition of bars 
and Hubble type, as defined by the family and stage axes of 
the RHS. 

  In a sense, this work represents a proof-of-concept study.
The current method is able to recognize many of the same 
structural properties of galaxies that have guided the study of 
galaxy morphology for the last fifty years. Having established 
how well the method works, under a variety of image \SN and 
resolution regimes, we can entertain the prospect of searching 
for refinements in the classification system itself. In other 
words, the next step is to apply various multivariate statistical 
techniques and non-supervised learning methods to determine if 
there are relationships among the Fourier-based model image 
parameters that convey more directly relationships among 
galaxies. For instance, strength of the $2\theta$ and 
$4\theta$ Fourier components in the inner regions of disk 
galaxies might be used to form a continuous measure of bar 
strength, and we might dispense with the 3-cell family 
axis of the RHS all together. Continuity along each axis 
of RHS was recognized long ago (de Vaucouleurs 1959), but 
human classifiers lacked the image measurement ability to 
classify galaxies in anything other than a set of bins. In 
addition, as discussed in Section 4.3, we may now incorporate 
information on image quality, as measured by \SN and resolution,
in determining what measurable morphological information 
is available in a galaxy image.  

 Finally, disk systems in the late stages of the RHS show increasing amounts 
of global image asymmetry. Such asymmetries are  sometimes linked to tidal 
encounters with nearby galaxies. How are these asymmetries related? Is the 
increased degree of disk asymmetry caused exclusively by encounters, and is 
the high frequency of asymmetric systems observed at $z$ \cge 0.5 the 
result of an increased interaction rate? Strength of the $1\theta$ 
Fourier component, as measured by the first principal component 
of the $1\theta$ amplitude profile, may be used to measure this
asymmetry independent of the presence of other morphological 
structures like bars and arms, something that is not possible 
with less sophisticated image-rotation asymmetry measurements. 
Such asymmetry measurements may be correlated closely with 
the internal kinematics of the galaxy (Rix \& Zaritsky 1996) 
or the present level of active star formation in a disk and 
hence will convey more direct insight into the physical processes 
driving galaxy formation and evolution.

  

We acknowledge support from NASA grants AR.6385.01.95A, 
GO-6609.01-95A, and AR-7534.01-96A.


\section*{REFERENCES} 
\begin{description} 

\item Abraham, R., Tanvir, N.R., Santiago, B., Ellis, R. S., 
      Glazebrook, K. G. \& van den Bergh, S., 1996, \mnras, 279, L47

\item Abraham, R., Valdes, F., Yee, H., \& van den Bergh, S. 1994,
      \apj, 432, 75 

\item Andreon, S., Gargiulo, G., Longo, G., Tagliaferri, R., Capuano, N.
2000, \mnras, 319, 700

\item Bazell, D. 2000, \mnras, 316, 519 

\item Bertin, E., \& Arnouts, K., 1997, \aap, 124, 163 

\item Bender, R. \& Mollenhoff, C. 1987, \aap, 177, 71

\item Buta, R. 1987, \apjs, 64, 383

\item Buta, R. \& Combes, F. 1996, Fund. Cosmic Physics, 17, 95

\item Buta, R., Mitra, S., de Vaucouleurs, G., \& Corwin, H.G. 1994,
      \aj, 107, 118.  

\item Cohen, S. H., Windhorst, R. A., Odewahn, S. C., Chiarenza, C. A. T.,
      \& Driver, S. P. 2001, \aj submitted 

Southern Galaxy Catalog (University of Texas Monograph in Astr., No.4) 

\item Corwin, H. G., de Vaucouleurs, A. \& de Vaucouleurs, G. 1985,
Southern Galaxy Catalog (University of Texas Monograph in Astr., No.4) 

\item de Vaucouleurs, G., de Vaucouleurs, A., Corwin, H.G., Buta, R., Paturel,
G. \& Fougu\'{e}, P. 1991, The Third Reference Catalog of Bright Galaxies,
Springer Verlag: New York (RC3)

\item Driver, S. P., Windhorst, R. A., \& Griffiths, R. E. 1995a, \apj, 453, 48   

\item Driver, S. P., Windhorst, R. A., Ostrander, E. J., Keel, W. C., 
   Griffiths, R. E., \& Ratnatunga, K.  U. 1995b, \apj, 449, L23      

\item de Vaucouleurs, G. 1977, {\it The Evolution of Galaxies and Stellar
   Populations}, ed. B. Tinsley and R. Larson, 43, (New Haven, USA:Yale
   University Press) 

\item de Vaucouleurs, G. 1958, Rev. Modern Phys., 30, 926 

\item de Vaucouleurs, G. 1959, {\it Handbuch der Physik}, 53, 311

\item de Jong, R. S., \& van der Kruit, P. C. 1994, \aaps, 106, 451.

\item Elmegreen, B., \& Elmegreen, D. 1985, \apj, 288, 438 

\item Frei, Z. \etal 1996, \aj, 111, 174 

\item Gal R.R., de Carvalho, R.R., Brunner, R., Odewahn, S.C., \& 
      Djorgovski, S.G., 2000, \aj, 120, 540 

\item Glazebrook, K., Ellis, R. E., Santiago, B., \& Griffiths, R. E. 1995, 
      \mnras, 275, L19       

\item Hickson, P. 1993, Astrophysical Letters and Communications, 29, 1 
 
\item Hubble, E. 1926, \apj, 64, 321

\item Kent, S. 1986, \aj, 91, 1301

\item Lauberts, A., and Valentijn, E.A., 1989, The Surface Photometry
Catalogue of the ESO-Uppsala Galaxies (European Southern Observatory, 
Garching)   

\item Lauer, T. 1985, \apjs, 57, 473

\item Mahonen, P. H. \& Hakala, P. J. 1995, \apjl, 452, L77

\item Mahonen, P.  \& Frantti, T. 20001 \apj, 541, 261 

\item Naim, A., Lahav, O., Sodre, L., \& Storrie-Lombardi, M. C. 1995,
      \mnras, 275, 567

\item Naim, A., Ratnatunga, K. U., Griffiths, R. E. 1997, 
      \apj, 476, 510  

\item McClelland, J. L. \& Rumelhardt, D. E. 1988, Explorations in 
Parallel Distributed Processing, (MIT Press), Cambridge, MA 

\item Nilson, P. 1973, Uppsala General Catalog of Galaxies, Roy. Soc. Sci.,
Uppsala (UGC)

\item Philip, N. S, Joseph, K. B., Kembhavi, A., \& Wadadekar, Y 2000, 
    Proceedings of Automated Data Analysis in Astronomy, Narosa Publishing, 
    New Delhi, 125

\item Rix, H. \& Zaritsky, D. 1995, \apj, 447, 82 

\item Roberts, M. S. \& Haynes, M. P. 1994, \araa, 32, 115

\item Odewahn, S.C. 1997, Nonlinear Signal and Image Analysis, 
      Annals of the New York Academy of Sciences, 188, 184 

\item Odewahn, S.C., Windhorst, R. A., Driver, S. P., \& Keel, W. C. 1996,
      \apjl, 472, L13

\item Odewahn,S.C., Stockwell, E.B., Pennington, R.M., Humphreys, R.M.,
      and Zumach, W.A.  1992, \aj, 103, 318

\item Odewahn, S. C, \& Aldering, G. 1995, \aj, 110, 2009

\item Odewahn, S. C. 1995, \pasp, 107, 770

\item Odewahn, S.C., Stockwell, E.B., Pennington, R.M., Humphreys, R.M.,  
\& Zumach, W.A.  1992, \aj, 103, 318 

\item Odewahn, S.C. 1989, Properties of the Magellanic Type Galaxies, 
       Ph.D. thesis, Univ. of Texas

\item Ohta, K, Hamabe, M., \& Wakamatsu, K. 1990, \apj, 357, 71 

\item Simien, F \& de Vaucouleurs, G. 1986, \apj, 302, 564 

\item Spiekermann, G., 1992, \aj, 103, 2102

\item Storrie-Lombardi, M.C., Lahav, O., Sodre, L.J., and
           Storrie-Lombardi, L.J. 1992, \mnras, 258, 8

\item van den Bergh, S., Cohen, J. G., \& Crabbe, C. 2001, \aj, 122, 611

\item Weir, N., Fayyad, M. F., \& Djorgovski, S. 1995, \aj, 109, 2401

\item Whitmore, B. C. 1984, \apj, 278, 61

\end{description} 

\clearpage


\clearpage 















\newpage 

\clearpage 

\clearpage 

\clearpage 

\clearpage 

\clearpage 

\clearpage 

\clearpage 

\clearpage 

\clearpage 

\clearpage 

\clearpage 

\begin{deluxetable}{c c c c c c c c}
\tablewidth{33pc}
\tablecaption{Results of Human Classifier Stage Comparisons }
\tablehead{
\colhead{$T_{x}$} &  \colhead{$T_{y}$}  & \colhead{S}  & \colhead{a}  &
\colhead{b}     & \colhead{R} & \colhead{N} & \colhead{$\sigma$}  } 
\startdata
SCO & RC3  & L & 1.00$\pm$0.05 &  0.022$\pm$0.148 & 0.898 & 119 & 1.60 \\
SHC & RC3  & L & 0.98$\pm$0.06 &  0.089$\pm$0.213 & 0.810 & 123 & 2.13 \\
RAW & RC3  & L & 1.02$\pm$0.05 & -0.060$\pm$0.159 & 0.889 & 121 & 1.61 \\
SHC & SCO  & L & 1.00$\pm$0.08 & -0.083$\pm$0.275 & 0.739 & 123 & 2.53 \\
RAW & SCO  & L & 1.00$\pm$0.04 &  0.060$\pm$0.118 & 0.929 & 120 & 1.38 \\
RAW & SHC  & L & 1.02$\pm$0.07 & -0.077$\pm$0.240 & 0.792 & 121 & 2.21 \\ 
    &      &   &               &                  &       &     &      \\
SCO & HST1 & D & 1.02$\pm$0.04 &  0.045$\pm$0.087 & 0.911 & 117 & 1.92 \\
SHC & HST1 & D & 0.99$\pm$0.07 & -0.011$\pm$0.161 & 0.786 & 125 & 2.86 \\
RAW & HST1 & D & 0.98$\pm$0.06 & -0.107$\pm$0.154 & 0.810 & 137 & 2.73 \\
SCO & SHC  & D & 0.99$\pm$0.07 &  0.113$\pm$0.148 & 0.795 & 116 & 2.84 \\
RAW & SCO  & D & 0.99$\pm$0.06 &  0.114$\pm$0.117 & 0.860 & 113 & 2.41 \\
RAW & SHC  & D & 0.99$\pm$0.08 & -0.046$\pm$0.201 & 0.729 & 123 & 3.27 \\
\enddata
\label{tab1}
\begin{flushleft}
The first two columns indicate the stage catalogs used on 
each axis to fit the relation $T_{y}=aT_{x}+b$. The column
labeled S indicates whether a sample of local (L) or 
distant (D) galaxies were used. The correlation coefficient, R,
the number of galaxies in the fit, N, and the standard 
deviation of the stage residuals, $\sigma$ are also tabulated. 
\end{flushleft}
\end{deluxetable}                

\begin{deluxetable}{l c c c c c}
\tablewidth{33pc}
\tablecaption{Accidental Stage Errors by Classifier }
\tablehead{
\colhead{Sample} &  \colhead{RC3}  & \colhead{HST1}      & \colhead{SCO}  &
\colhead{RAW}     & \colhead{SHC} } 
\startdata
Local & $0.97\pm0.35$  & - & $1.26\pm0.26$ & $0.88\pm0.39$ & $2.02\pm0.16$ \\
Distant & - & $1.53\pm0.13$ & $1.20\pm0.13$ & $2.16\pm0.09$ & $2.48\pm0.08$ \\
\enddata
\label{tab2}
\end{deluxetable}                

\begin{deluxetable}{l l c}
\tablewidth{20pc}
\tablecaption{The Palomar Galaxy Sample}
\tablehead{
\colhead{name} &  \colhead{RC3 Type}  & \colhead{filters} } 
\startdata
   NGC 1744  &  SB(s)d    &  1g      \\
   NGC 4501  &  SA(rs)b   &  2B      \\
   NGC 4519  &  SB(rs)d   &  1g      \\
   NGC 4618  &  SBm       &  2g      \\
   NGC 5850  &  SB(r)b    &  1g,1B   \\
   NGC 5985  &  SAB(r)b   &  1g      \\
   NGC 5377  &  SB(s)a    &  1g      \\
   NGC 6070  &  SA(s)c    &  3g,1B   \\
   NGC 6340  &  SA(s)0    &  1g      \\
   NGC 6384  &  SAB(r)bc  &  1g,2B   \\
   NGC 6643  &  SA(rs)c   &  1g      \\
   NGC 6764  &  SB(s)bc   &  1g      \\
   NGC 6824  &  SA(s)b    &  1g      \\
   NGC 6951  &  SAB(rs)bc &  2g      \\
   NGC 7217  &  SA(r)ab   &  2g      \\
   NGC 7457  &  SA(rs)0   &  1g      \\
   NGC 7479  &  SB(s)c    &  1g      \\
   NGC 7741  &  SB(rs)cd  &  1g      \\
   PGC 61512  &  SB(s)dm   &  1g      \\             
\enddata
\label{P60table}
\end{deluxetable}                

\begin{deluxetable}{l c l}
\tablewidth{30pc}
\tablecaption{LMORPHO Image Parameters Used for Classification} 
\tablehead{
\colhead{name} &  \colhead{Feature Set}  & \colhead{Comment} } 
\startdata
PC1\_Fnorm &  1,2,3   &  First PC of normalized flux profile   \\
PC2\_Fnorm &  1,2,3   &  Second PC of normalized flux profile  \\
PC1\_1TA   &  2,3     &  First PC of $1\theta$ amplitude profile \\
PC1\_2TA   &  1,2,3,4 &  First PC of $2\theta$ amplitude profile \\ 
PC2\_2TA   &  1       &  Second PC of $2\theta$ amplitude profile \\ 
PC1\_4TA   &  1,2     &  First PC of $4\theta$ amplitude profile \\ 
PC2\_4TA   &  1,2,4   &  Second PC of $4\theta$ amplitude profile \\ 
PC1\_4TP   &  1,2     &  First PC of $4\theta$ phase profile \\ 
PC2\_4TP   &  1,2,4   &  Second PC of $4\theta$ phase profile \\ 
S/N        &  3       & Mean \SN measured in elliptical aperture \\
$N_{pix}$  &  3       & Number of image pixels \\
$B_{P3}$   &  4       & Bar parameter 3 \\
$B_{P4}$   &  4       & Bar parameter 4 \\
\enddata
\label{ParList}
\end{deluxetable}                

\begin{deluxetable}{c l c c}
\tablewidth{30pc}
\tablecaption{Train/Test Galaxies Divided by Stage}
\tablehead{
\colhead{node number} & \colhead{Number}  & \colhead{T interval}  & \colhead{Literal range} } 
\startdata
  1   &  19  & -8.0 -4.5  & E to E0-1  \\
  2   &   6  & -4.5 -2.5  & E+ to L0-  \\
  3   &  13  & -2.0 -0.5  & L0 to L0+  \\
  4   &  18  & -0.5 1.5   & S0a to Sa  \\
  5   &  34  & 1.5  3.5   & Sab to Sb  \\
  6   &  91  & 3.5  5.5   & Sbc to Sc  \\
  7   &  30  & 5.5  7.5   & Scd to Sd  \\
  8   &  12  & 7.5  10.5  & Sdm to Im  \\
\enddata
\label{Tbreakdown}
\end{deluxetable}                
 
\begin{deluxetable}{l l l l l l c c c}
\tablewidth{30pc}
\tablecaption{Results for Backpropagation ANN Stage Classifiers}
\tablehead{
\colhead{$\alpha^{TR}$} & \colhead{$\sigma_{1}^{TR}$}  & \colhead{$\sigma_{2}^{TR}$}  & \colhead{$\alpha^{TE}$} & \colhead{$\sigma_{1}^{TE}$}  & \colhead{$\sigma_{2}^{TE}$} & \colhead{Epoch} & \colhead{Feature Set} & \colhead{l}  } 
\startdata
  0.920 & 0.93 & 2.09 & 0.908 & 2.25 & 2.09 &  5800 & 1 &  9  \\
  0.953 & 1.15 & 1.71 & 0.968 & 2.29 & 1.71 &  2900 & 1 & 11  \\
  1.045 & 1.16 & 1.68 & 0.949 & 2.46 & 1.68 &  2600 & 1 & 12 $\ast$ \\
  0.935 & 1.20 & 1.84 & 0.959 & 2.37 & 1.84 &  2000 & 1 & 13  \\
  0.705 & 1.71 & 2.57 & 0.722 & 1.92 & 2.57 &  1000 & 1 & 14  \\
  0.931 & 1.78 & 2.28 & 0.906 & 1.92 & 2.28 &  4200 & 2 &  9  \\
  0.974 & 1.35 & 1.82 & 0.909 & 2.06 & 1.82 &  4300 & 2 & 11  \\
  0.936 & 1.56 & 2.25 & 0.914 & 2.06 & 2.25 &  2300 & 2 & 12  \\
  0.938 & 1.13 & 1.87 & 0.986 & 2.51 & 1.87 &  3900 & 2 & 13 $\ast$ \\
  1.081 & 1.32 & 1.74 & 1.005 & 1.88 & 1.74 &  3600 & 2 & 14  \\
  1.009 & 1.44 & 1.71 & 0.978 & 1.71 & 1.71 &  1800 & 3 & 9   \\
  0.878 & 1.16 & 2.59 & 0.803 & 1.77 & 2.59 &  1000 & 3 & 11  \\
  1.016 & 0.93 & 1.54 & 0.946 & 1.83 & 1.54 &   900 & 3 & 12 $\ast$ \\
  0.917 & 1.36 & 2.07 & 0.965 & 1.62 & 2.07 &   500 & 3 & 13  \\
  1.025 & 0.79 & 1.38 & 1.045 & 1.79 & 1.38 &   900 & 3 & 14  \\
\enddata
\label{StageClassStats}
\begin{flushleft}
Asterisks indicate networks that gave the best performance for a
given node architecture and these were used to compute mean types in 
Figure~\ref{net_results}. The TR superscript refers to training 
results and TE refers to testing results. The l column indicates 
the number of nodes in each of the two hidden layers used in 
every ANN. As explained in the text, the Feature Set number 
indicates the set of image parameters fed to each network.  
\end{flushleft} 
\end{deluxetable}                

\end{document}